\documentclass[article,useAMS,usenatbib]{mn2e}
\usepackage{graphicx}
\usepackage{eufrak}
\usepackage{eucal}
\usepackage{txfonts}

\begin{document}

\title[From flux to dust mass]{From flux to dust mass: Does the grain-temperature distribution matter for estimates of cold dust masses in supernova remnants?}

\author[Mattsson et al.]{Lars Mattsson$^{1,2}$\thanks{E-mail: larsmat@kth.se},  Haley L. Gomez$^3$, Anja C. Andersen$^1$ \& Mikako Matsuura$^4$\\
$^1$Dark Cosmology Centre, Niels Bohr Institute, University of Copenhagen, Juliane Maries Vej 30, DK-2100, Copenhagen \O, Denmark\\
$^2$Nordita, KTH Royal Institute of Technology \& Stockholm University, Roslagstullsbacken 23, SE-106 91, Stockholm, Sweden\\
$^3$School of Physics \& Astronomy, Cardiff University, Queens Buildings, The Parade, Cardiff, CF24 3AA, UK\\
$^4$Department of Physics \& Astronomy, University College London, Gower Street, London WC1E 6BT, UK}

\pagerange{\pageref{firstpage}--\pageref{lastpage}} \pubyear{2013}

\maketitle

\label{firstpage}

\date{\today}

\begin{abstract} The amount of dust estimated from infrared to sub-millimetre (submm) observations strongly depends on assumptions of different grain sizes, compositions and optical properties. Here we use a simple model of thermal emission from cold silicate/carbon dust at a range of dust grain temperatures and fit the spectral energy distribution (SED) of the Crab Nebula as a test. 
This can lower the derived dust mass for the Crab by $\sim$50\% and 30-40\% for astronomical silicates and amorphous carbon grains compared to recently published values ($0.25M_{\sun} \to 0.12M_{\sun}$ and $0.12M_{\sun}\to 0.072M_{\sun}$, respectively), but the implied dust mass can also increase by as much as almost a factor of six ($0.25M_{\sun} \to 1.14M_{\sun}$ and $0.12M_{\sun}\to 0.71M_{\sun}$) depending on assumptions regarding the sizes/temperatures of the coldest grains. The latter values are clearly unrealistic due to the expected metal budget, though. Furthermore, we show by a simple numerical experiment that if a cold-dust component does have a grain-temperature distribution, it is almost unavoidable that a two-temperature fit will yield an incorrect dust mass estimate. But we conclude that grain temperatures is not a greater uncertainty than the often poorly constrained emissivities (i.e., material properties) of cosmic dust, although there is clearly a need for improved dust emission models. The greatest complication associated with deriving dust masses still arises in the uncertainty in the dust composition.
\end{abstract}


\section{Introduction}
Observations suggest Ly$\alpha$ systems, quasars and gamma-ray burst hosts at high-$z$ contain very large amounts of dust \citep[see, e.g.][]{Bertoldi03, Beelen06, Michalowski08, Michalowski10a,Michalowski10b}, which forces models of stellar dust production and galactic dust evolution to extremes in order to reproduce these results \citep{Morgan03,Dwek07,Gall11a,Gall11b,Mattsson11}. Assuming stars as the primary source of dust suggests the dust is due to  supernovae (SNe) and a top-heavy initial mass function favouring the formation of SNe. This scenario can in principle explain the observations, provided the dust-destruction rate in the interstellar medium is {\it much} lower than expected based on the kinetics of SN explosions \citep[see][]{Morgan03,Dwek07,Gall11a,Gall11b,Jones11,Mattsson11}.

Recent observations in the FIR/sub-mm of nearby core-collapse supernova remnants (SNRs), with ages $> 25$ years, suggest high dust-formation efficiencies in the SN ejecta \citep[][henceforth G12]{Barlow10, Matsuura11, Gomez12} though these are limited in sample size.  The dust masses estimated using canonical single or two-temperature component fits to the spectral energy distributions in the FIR/-submm imply dust masses which are uncomfortably close to (or indeed exceed) the amount of metals predicted to be ejected in these supernovae \citep[see e.g.][G12]{Matsuura11}. The Inferred dust mass depends on the assumed dust properties, which can thus change whether SNRs appears to contain unrealistically large amounts of dust or not. If the grains are silicates, the grain masses are high relative to the emissivity. For carbonaceous dust the mass/emissivity-ratio is much smaller, which is also the case for ice-coated grains and various other types of coagulates.
 
Variation in dust temperatures, grain sizes and material properties makes dust-mass estimates from SEDs uncertain, which is recognised in the literature. The fact that SEDs arising from dust emission often suggest there is a range of grain temperatures in a dust population, has inspired several publications on how to solve the inverse problem of finding the corresponding grain-temperature distribution (GTD) from an observed SED \citep[see, e.g.,][]{Xie91,Xie93,Li99}. Solving this inverse problem would thus provide empirical constraints on the functional form of the GTD that can be used to improve dust estimates. In principle, there is a detailed correspondence between the dust SED and the GTD, but finding the exact shape of the latter is unfortunately often a badly conditioned problem \citep{Hobson94}. Common practice is still to fit one or two single-temperature components to model the SED, which may affect the implied dust mass; in particular, the mass of cold dust ($\sim20$\,K) may be overestimated by pushing too much flux into this component.

It has recently been suggested by, e.g., G12, \citet{Richardson13} and \citet[][henceforth TD13]{Temim13} that the treatment of grain temperatures may be important in the case of the Crab Nebula.  Initially G12 used a two-temperature fit to the observed SED to derive dust masses of $0.25\,M_{\sun}$ and $0.12\,M_{\sun}$ for silicates and carbon within the Crab respectively.  In parallel with our work, TD13 have made a model of the FIR dust emission in the Crab Nebula explicitly taking the synchrotron radiation field and grain-sizes into account in calculating the radiation balance. They claim that the required dust mass to explain the SED is significantly reduced in their model but this is due to a combination of the model as such and the use of a different set of optical constants (for which laboratory measurements are only available at $\lambda < 300\,\mu$m). We note, also, that detailed modelling can give very different results compared to TD13 \citep{Owen15}. It is therefore worth elaborating on the effect of considering a cold-dust GTD and in particular, to consider how sensitive the dust masses obtained from SED fitting are to the assumed dust temperatures.  We use a simple, computationally inexpensive, `model' based on known constraints on the GTD, which can be applied in any environment regardless of whether the radiation field heating the dust can be specified or not.  \citet{Kovacs10, Magnelli12}, have successfully modelled the GTD of star forming galaxies using a power-law in a similar manner.  Attempts to solve the inverse problem mentioned above for molecular clouds have suggested the GTD may be closer to an exponential form, however \citep[see, e.g.,][]{Xie93}.  The purpose of this paper is not to provide a precise and realistic model of dust emission, though. We merely aim to investigate how much of a difference introducing a range of grain temperatures makes in general, compared to other uncertainties in deriving the dust mass.

\section{FIR/sub-mm emission from dust}
In this section we will just briefly summarise the physics underlying the conventional way of inferring dust masses and grain temperatures from FIR/sub-mm fluxes\footnote{ In the present paper we refer to `flux' in the following way: the total flux $F$ from a source is the energy output (luminosity) per unit surface area, e.g. in units of W~m$^{-2}$. The {\it specific} or {\it monochromatic} flux is per  wavelength (or frequency) $F_\lambda$ is the same as `spectral flux density' commonly used in observational work (usually with the unit being Jy~$ = 10^{-26}$~W~m$^{-2}$~Hz$^{-1}$). However, to keep the terminology simple we only use the term `flux', and ask the reader to bear in mind that a subscript $\lambda$ means the flux at a specific wave length.}.

\subsection{Dust masses}
It is normally assumed that dust grains absorb and emit photons according to Kirchhoff's law and that the source function can be described by a Planck function $B_\lambda\,|d\lambda/d\nu|$ and that optically thin conditions apply in the FIR/sub-mm of the surrounding medium. If $Q_{\rm em} = Q_{\rm abs}$, where $Q$ denotes the ratio of the effective and geometric cross sections (Kirchhoff's law) and $\rho \kappa_\lambda \Delta x\ll  1$, where $\rho$ is the gas density, $\kappa_\lambda$ is the opacity due to absorption and $\Delta x$ is the geometric thickness of the surrounding medium (optically thin conditions), then the observed FIR/sub-mm flux is the sum of all reemitted flux from dust grains, where the flux from a grain of radius $a$ and temperature $T_{\rm d}$ is  
 \begin {equation}
F_\lambda^{\rm gr}(a) = \pi\,D^{-2}\,B_\lambda(T_{\rm d})\,a^2 Q_{\rm abs}(\lambda, a)\, n(a)\,da.
\end{equation}
In the equation above, $D$ is the distance to the observer and $n(a)$ is the grain-size distribution (GSD) by number normalised per unit volume. With the additional assumption that an ensemble of dust grains can be described with a single grain temperature $T_{\rm d}$ and that $a/\lambda \ll 1$ (the Rayleigh limit), the flux from this ensemble can be expressed  as
\begin{equation}
F_\lambda^{\rm dust} = {3\over4}{B_\lambda(T_{\rm d})\,Q'_{\rm abs} V_{\rm gr}\over D^2}, 
\end{equation}
where $Q'_{\rm abs}(\lambda) = Q_{\rm abs}(\lambda, a)/a$ in the Rayleigh limit and $V_{\rm gr}$ is the total volume taken up by the dust. Multiplying both sides of the above equation with the bulk density of the grains $\rho_{\rm gr}$, one obtains the relation \citep{Hildebrand83,Gall11}
 \begin {equation}
 \label{dustmass}
M_{\rm d} = {4\over 3} {\rho_{\rm gr}\,D^2 \,F_\lambda^{\rm dust}\over Q'_{\rm abs}\, B_\lambda(T_{\rm d})} = {D^2 \,F_\lambda^{\rm dust}\over \tilde{\kappa}_\lambda\,B_\lambda(T_{\rm d})},
\end{equation}
where $\tilde{\kappa}_\lambda$ is a quantity to be referred to as  `emissivity' (or `absorptivity', depending on the context) which has the same dimension as opacity, but reflects the optical properties of the grain material and should not be confused with the opacity of the surrounding medium $\kappa_\lambda$ referred to above.

Eq. (\ref{dustmass}) conveniently evades the essentially unknown GSD. However, it is not obvious that the assumptions made to derive this simple relation between dust flux and mass are valid in general. In particular, it is only valid in the Rayleigh limit and if all dust grains have the same temperatures, where the latter implies that they should all have the same size as well. The Rayleigh limit only applies to emission at long wavelengths (absorption and scattering is mainly in the optical/UV), which may cause an implicit dependence on grain size with different grain temperatures as the result. In any realistic dust model one has to integrate over a suitable size distribution and make use of a variety of dust compositions/materials. That is, the dust mass corresponding to a given SED depends on the GTD $W(T_{\rm d}) \equiv dM_{\rm d}/dT_{\rm d}$. The flux from dust for such a distribution is given by
 \begin {equation}
 \label{dustmass2}
F_\lambda^{\rm dust} = \tilde{\kappa}_\lambda D^{-2}\int_{T_{\rm low}}^{T_{\rm high}} W(T_{\rm d}) B_\lambda(T_{\rm d})\,dT_{\rm d}.
\end{equation}
We will in the following elaborate on how much of an effect $W(T_{\rm d})$ has on the dust masses derived from FIR/-sub-mm SEDs.

\subsection{Grain temperatures}
The basic assumption underlying the relation between dust flux and dust mass is that it is the same dust grains that absorb light in the UV/optical or gain energy due to collisions with other particles (usually electrons) that are also emitting radiation in the IR/FIR/sub-mm. The physical underpinnings for radiative and collisional heating are different, however, but the equilibrium temperature distributions for the grains may not be that different.

\subsubsection{Radiative heating and cooling}
\label{radheat}
Radiative heating is normally due to absorption of radiation in the UV/optical and assuming the grains are in local thermal equilibrium with the mean intensity of the radiation field we can equate the absorbed and emitted power, i.e. $P_{\rm abs} = P_{\rm em}$. The optical depth of the surrounding medium may in principle affect the energy absorption but has little effect on the re-emission at long wavelengths. Thus, making use of $Q'_{\rm abs}$ (as previously defined) in the Rayleigh limit and assuming the 
surrounding medium is optically thin, we have
\begin{equation}
\label{ebalance1}
 \int_0^\infty Q_{\rm abs}(\lambda,a)\,{J}_{\star, \lambda} \,d\lambda \approx 4\pi a \int_0^\infty Q'_{\rm abs}(\lambda) B_{\lambda}(T_{\rm d}) \,d\lambda,
\end{equation}
where ${J}_{\star, \lambda}$ is the mean intensity of the radiation field. Note that $Q'_{\rm abs}$ (the Rayleigh-limit approximation) can only be used on the right-hand side of Eq. (\ref{ebalance1}), which represents emission at long wavelengths,  and that Eq. (\ref{ebalance1}) is strictly valid only locally (but can be generalised, resulting in an effective overall GTD). Conveniently, the emission at long wavelengths is approximately a power-law in $\lambda$. Thus, replacing $Q'_{\rm abs}$ with $Q'_0 (\lambda/\lambda_0)^{-\beta}$, integrating over wavelength and assuming that heating occurs in the grey-absorption limit  (or large-particle limit) where $Q_{\rm abs}$ is constant, we have then a simple power-law of the form 
\begin{equation}
\label{powerlaw1}
T_{\rm d}(a) =  T_{\rm s} \left({a\over a_{\rm s}}\right)^{-1/(4+\beta)},
\end{equation}
where $T_{\rm s}$ and $a_{\rm s}$ are scaling parameters. Note that even if one adopts an explicitly grain-size dependent $Q_{\rm abs}$, the grain temperature is still (locally) uniquely determined by the grain size under thermal equilibrium conditions. For further details on the above, see Appendix \ref{hecoeq}.

The GTD $W(T_{\rm d})$ can be determined from
\begin{equation}
\label{powerlaw2}
W(T_{\rm d}) = \varphi(a)\left|{dT_{\rm d}\over da}\right|^{-1} ,
\end{equation}
where $\varphi(a)$ is the GSD in terms of mass. For a canonical MRN  distribution \citep{Mathis77} $\varphi(a) \propto a^{-0.5}$. Thus, adopting an MRN distribution and the grey-absorption limit,  the GTD is simply also a power-law $W(T_{\rm d}) \propto  T_{\rm d}^{-3-\beta/2}$. Such a GTD has been recovered for hot dust (heated by short-wavelength radiation) around active galactic nuclei \citep{Wang96}. The grey-absorption limit may not be strictly applicable in other cases (e.g., SNRs) and the slope of the temperature distribution is therefore likely steeper than in Eq. (\ref{powerlaw2}) and may deviate from the simple power-law form above, i.e., the scale temperature $T_{\rm s}$ would in this case be a function of $a$. We note also that collisional heating should give rise to a GTD in a way similar to radiative heating (see Appendix \ref{collheat}).

\subsubsection{Temperature fluctuations?}
\label{fluct}
%Previous studies \citep[e.g.,][]{Dwek86} have demonstrated theoretically how the GTD is affected for small grains undergoing significant temperature fluctuations as a result of collisional heating: the GTD becomes wider and tends towards a power-law shape for really small grains. A corresponding phenomenon can be found for radiative heating too, since temperature fluctuations in small grains due to absorption of energetic photons is quite similar \citep{Draine03}.
Previous work has shown how the GTD is affected for small grains undergoing significant temperature fluctuations (non-equilibrium conditions) as a result of being hit by energetic photons or other particles \citep[see, e.g.,][and references therein]{Purcell76,Aannestad79,Draine85,Dwek86,Draine01,Draine03}. The heat capacity determines how much energy the dust particle can hold, i.e., the energy in an absorbed photon will partly heat up the material and the remainder will be emitted at longer wavelengths. The amount of energy a grain can hold is proportional to its mass, which is why small grains do not easily obtain an equilibrium temperature if heated with short-wavelength radiation. For radiative heating, we have that an energy increment $dE$ is causing a temperature increment according to $dE = C_V\, dT_{\rm d}$, where $C_V$ is the heat capacity at a constant volume. The heat capacity for a spherical grain is, in the low-temperature limit of Debye's model ($T_{\rm d}/T_{\rm D}\ll 1$, with $T_{\rm D}$ the Debye temperature),
\begin{equation}
\label{debye}
C_V \approx {12\pi^4\over5} \,kN\, \left({T\over T_D}\right)^3 = {16\pi^5\over5}{k\,\rho_{\rm gr} a^3\over m_{\rm X}} \left({T_{\rm d}\over T_{\rm D}}\right)^3,
\end{equation}
where $N$ is the number of monomers in the grain, $m_{\rm X}$ their mass \citep{Aannestad79,Draine85}. The temperature increase is effectively $\Delta T_{\rm d} \sim Q_{\rm abs}(a,\lambda)\, C_V^{-1}\,E_{\rm phot}\sim E_{\rm phot}^{1/4}$, where $E_{\rm phot}$ is the energy of the photon and $Q_{\rm abs}$ is the ratio of the effective and geometrical cross sections of the grain. Inserting suitable numbers shows that a single UV photon could suffice to raise the temperature of a cold ($\sim 30\,$K) nano-sized grain almost an order of magnitude. Very small grains will usually also cool rapidly and regain their initial temperatures of $\sim 30\,$K in typically a couple of hours or less \citep{Purcell76,Draine01}.  Thus, at any given time, a population of cold, very small grains may therefore have wide range of temperatures, leading to a GTD even for a population of grains with exactly the same sizes. 

At first glance it may seem like the above factors complicate the picture a lot. But, fortunately, the effect on the estimated mass of cold dust is negligible. First, most of the mass is expected to be in the large grains (cf. the MRN size distribution in terms of dust mass, $d\rho_{\rm d}/da \equiv \varphi(a) \propto a^{-0.5}$) and dust in SNe appears to have a strong bias towards large grains in general according to recent results \citep{Gall14,Wesson14,Owen15}. Not more than a few percent of the mass can be in grains small enough to undergo significant temperature fluctuations \citep{Purcell76,Aannestad79}. From eq. (\ref{debye}) we can also see that size can easily compensate for the fact that the grains are cold. Second, if temperature fluctuates are important, they will cause emission in the infrared and near infrared bands \citep{Li01}. Cold dust radiates almost exclusively in the FIR/sbmm bands. Thus, in conclusion, we do not have to worry greatly about temperature fluctuations as long as we are dealing with cold dust and a steep GTD.

\section{The effects a grain-temperature distribution}
\label{multitemp}
Under most circumstances it is inevitable that there is a distribution of dust temperatures rather than distinct representative dust temperatures for specific dust components (e.g., a cold, warm and hot component) as we have touched upon above. This fact has previously been pointed out as a caveat \citep[see, e.g., G12, TD13 and][in the case of the Crab Nebula]{Richardson13}. Such a distribution must be very steep and narrow not to create SEDs which are inconsistent with the featureless `bumps' associated with cold dust, but it can still have a significant effect on the derived dust mass. In this Section we describe a simple multi-temperature model for the FIR/sub-mm dust emission. We use the Crab Nebula as a test case, { to show that an observed SED can be reproduced with very different dust masses}, and then continue with a more general analysis of how multi-temperature SED fits may differ from two-temperature fits, by generating a large mock sample of SEDs from simple GTDs.

\subsection{Grain-size or temperature distribution?}
There are essentially just two ways to incorporate the effect of a range of grain temperatures when modelling an SED: either one has to (computationally) find the GTD from an assumed GSD and a (possibly) known radiation field, or, one could make a direct assumption or estimate of the GTD (or its functional form, more precisely). TD13  preferred the first approach, while we will here explore the latter. 

There is no actual advantage of any of these two approaches over the other in terms of their physical correctness (we simply have too little information to distinguish between the two), though it is clear that an assumption directly regarding the GTD is much simpler to deal with. There is also a direct link between the observed SED and the corresponding GTD that could (in theory) be used to constrain the functional form \citep{Xie91,Xie93,Li99}. In principle one could thus construct a fairly consistent model even without any knowledge about the heating radiation field. 

%TD13 assumed the GSD is a power-law with the power index and the upper size limit as free parameters (the lower size limit fixed at $a_{\rm min}= 0.001\,\mu$m). The corresponding temperature distribution is not a simple power-law or exponential, but since we do not know exactly what form it should have, it is not obviously any more or less correct than the model proposed here. If, on the other hand, one assumes a specific functional form of the GTD instead (as we aim to do here) there is no need to make any further assumptions about the GSD to obtain a dust mass, since all that is needed is the GTD. Any such ansatz may correspond to a GSD that is not necessarily a power-law, but the exact functional form of the GSD of SN dust is not known either.

One could of course argue that the best option would be to use a GSD that results from theoretical modelling of SN dust formation and then chose an approach similar to that used by TD13. But unfortunately the models do not offer a consistent picture. \citet{Nozawa03} suggest the GSD produced by SNe may be somewhat flatter [$n(a) \propto a^{-2.5}$], except  for very large grains, where the MRN distribution seems to be recovered. But the effective GSD may be even flatter still, if small grains are destroyed \citep{Nozawa07}, and the upper and lower size limits do not correspond to sharp cut-offs, so the assumption of a power-law GSD made by TD13 is not obviously the optimal ansatz. 

\subsection{Multi-temperature SED}
Modelling of the dust contribution to the SED is often done by fitting one or two components (cold and warm dust),
with weighting factors that specify the mass contribution from each component. We generalise this to $N$ components, but with the weighting factors constrained by a GTD assumed to follow a power law or exponential as motivated above. Thus, we maintain the same number of parameters (four) when fitting the SED, while at the same time dust grains can have an arbitrary range of temperatures. 

Regarding the functional form of the GTD $W(T_{\rm d})$, the energy balance (Eq. \ref{ebalance1}) suggest that if a grain has a temperature $T_1$ and another grain has a temperature $T_2>T_1$, then grains with temperature $T_2$ cannot be larger than grains with temperature $T_1$ (see also the Appendix \ref{hecoeq}). That is, $dW/dT_{\rm d}\leq 0$ for all $T_{\rm d}$. Furthermore, because $Q_{\rm abs}$, at any given wavelength, normally is a smooth function of $a$ without very prominent features, there is no reason to expect that $W(T_{\rm d})$ is a complicated function of $T_{\rm d}$. It is therefore reasonable to assume that the GTD is a smooth, monotonously decreasing function of grain temperature. Steep power-laws have previously been considered in other contexts  \citep{Aguirre03,Kovacs10,Magnelli12}.

\subsubsection{Power-law}
As discussed previously, radiative heating that occurs in the grey-absorption limit corresponds to a power-law GTD $W(T_{\rm d})$  where the GSD is a power-law. If the dust is heated by short-wavelength radiation (UV, X-ray etc.) this is in fact a good approximation of reality. We have therefore as good reasons to try a power-law form as our ansatz for the effective GTD as there are reasons to expect a power-law GSD. 

With $W(T_{\rm d}) = W_0\,(T_{\rm d}/T_0)^{-\alpha}$, where we set $T_0 = T_{\rm low}$, we can model the flux from an ensemble of dust grains with different sizes and temperature using a function
\begin{equation}
\label{fitfunc}
S_\lambda(T_{\rm high},T_{\rm low};\alpha)  = S_0(N)\, \sum_{i=0}^{N-1}  \left({T_i\over T_{\rm low}} \right)^{-\alpha} B_\lambda(T_i) ,
\end{equation}
in which $T_i = i\,N^{-1}(T_{\rm high} - T_{\rm low}) + T_{\rm low}$, where $T_{\rm high}$ and $T_{\rm low}$ are the high and low temperature cut-off, respectively. $S_0$ is a constant such that $S_{\lambda}$ represents the blackbody flux from all $N$ components. This function replaces the Planck functions and the weighting coefficients in, e.g., a two-component model and the exponent $\alpha$ can be treated as a free parameter. If $N$ is large (we assume $N=100000$ in subsequent applications), the fitting result corresponds to a continuous GTD $W(T_{\rm d})$ as in Eq. (\ref{dustmass2}).

\subsubsection{Exponential}
The power-law form of $W(T_{\rm d})$ is strictly valid only under very special circumstances, as we have pointed out in Section \ref{radheat}. It is more realistic that the slope of the GTD changes with the grain temperature $T_{\rm d}$. The GTDs derived by \citet{Xie93} from the SEDs of cold molecular clouds are very close to exponential for $T_{\rm d} \gtrsim 25$\,K and detailed computer simulations of molecular clouds do in fact suggest this may be a realistic functional form of the GTD of a single dust species \citep{Bethell04}. If an exponential form of $W(T_{\rm d})$ works well for molecular clouds it seems reasonable it could work as well for cold dust in SNRs. Thus, we try also an exponential model. Using an exponential GTD, $W(T_{\rm d}) = W_0\exp(-T_{\rm d}/T_0)$, we have in analogy with the power law case,
\begin{equation}
\label{fitfunc2}
S_\lambda(T_{\rm high},T_{\rm low};T_0)  = S_0(N)\, \sum_{i=0}^{N-1}  \exp\left(-{T_i\over T_0} \right) B_\lambda(T_i), 
\end{equation}
where $T_0$ is a free scaling parameter. Since a GTD must be steep, a realistic value of $T_0$ is only a few K. 

\begin{figure}
  \resizebox{\hsize}{!}{
   \includegraphics{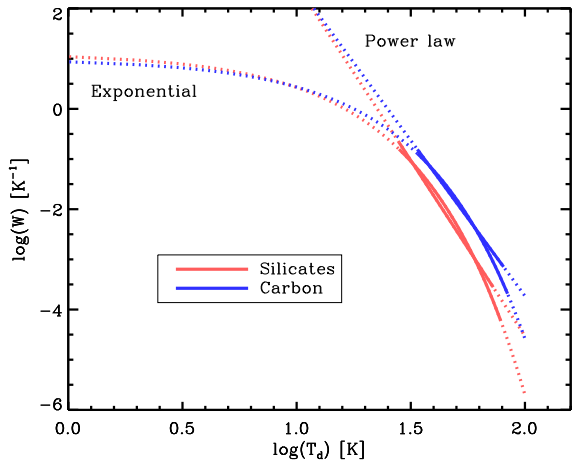}}
  \caption{\label{graintemp} Grain temperature distributions for the dust SED models with $T_{\rm low}$ fixed to the temperature of the cold component in a two-temperature fit of the Crab Nebula. The temperature interval used to model the SED for each model is shown by solid lines. See also Table \ref{parameters}.}
  \end{figure}

%To our knowledge, an exponential GTD has not been explicitly considered in previous studies where dust SEDs are modelled using GTDs of a specific functional form. But 
Power-law models used on very cold dust suggest the GTD is flattening out at low dust temperatures \citep{Aguirre03}, which further motivates that we should consider a GTD of the exponential form above (see Fig. \ref{graintemp} for an example of how this flattening naturally occurs at low $T_{\rm d}$ for an exponential GTD). 

\subsubsection{Temperature cut-offs}
Obviously, there has to be a lowest and a highest possible grain temperature, because grains are neither arbitrarily small, nor arbitrarily large (see Appendix \ref{hecoeq}). We use these cuts on the temperature distribution as fitting parameters as well (see Eq. \ref{fitfunc}), which corresponds to the integration limits above. Initial values for $T_{\rm low}$ and $T_{\rm high}$ may be determined by a two-temperature fit, which stabilises the fitting procedure. One may argue that the lower limit $T_{\rm low}$ may also be fixed to this value, since the lowest possible dust temperature is essentially defined by the shape of the FIR/-sub-mm tail of the SED (see fig. 4  in G12). But this is not necessarily a good approach, since the flux contribution from very cold (and large) grains may be small, while their contribution to the dust mass is significant. Note, also, that this is related to the upper grain-size limit as well (the coldest grains are the largest ones), while the upper temperature limit, and thus also the lower grain-size limit, is harder to constrain (we will return to this later).  TD13 assume an upper grain-size limit which, according to their model, corresponds to approximately the temperature of the cold component in a two-temperature fit. Therefore, in our `test application' to the Crab Nebula below, we have included a case where $T_{\rm low}$ is fixed to the temperature of the cold component in the two-temperature fit, in order to compare with the results of TD13.

    \begin{table*}
  \begin{center}
  \caption{\label{parameters} Comparison of parameters derived from fitting the SED of the Crab Nebula with two-temperature and GTD models. $T_{\rm low}$ and $T_{\rm high}$ denote the low and high cut offs of the GTD as well as the temperatures of the cold and warm components in the two-component models.  $\alpha$ is the resultant power-law index for the GTD. }
  \begin{tabular}{llllllllll}
  \hline
  \hline
Model & Dust type & $\rho_{\rm gr}$ & $T_{\rm low}$ & $T_{\rm high}$ & $T_0$ & $\alpha$ & $M_{\rm d}$  & $\chi^2$ & Remark\\   
            &    & [g/cm$^{3}$]     & [K]                   & [K] &  [K] & & [$M_{\sun}$]   &                 &      \\
    \hline
A &Astron. silicate & 3.3 & $28.1$ & $55.6$ & - & -     &  0.25 & 8.54 & Two-temperature fit.\\ 
B &Astron. silicate & 3.3 & $28.1$ & $72.6$ & - & 7.0 &  0.14 & 7.98 & Power-law GTD, $T_{\rm low}$ fixed.\\ 
C &Astron. silicate & 3.3 & $17.4$ & $68.7$ & - & 5.2 &  0.51 & 7.98 & Power-law GTD, $T_{\rm low}$ free.\\ 
D &Astron. silicate & 3.3 & $28.1$ & $78.3$ & 6.38 & - &  0.12 & 8.23 & Exponential GTD, $T_{\rm low}$ fixed.\\ 
E &Astron. silicate & 3.3 & ${\it 2.73}$ & $74.1$ & 6.37 & - &  1.14 & 7.87 & Exponential GTD, $T_{\rm low}$ free.\\ [1mm]

F &Amorphous carbon & 1.81 & $33.8$ & $63.4$ & - & -     &  0.12 & 8.19 & Two-temperature fit.\\
%G &Amorphous carbon & 1.81 & $33.8$ & $80.6$ & - & 6.2 &  0.085 & 7.46 & Power-law GTD, $T_{\rm low}$ fixed.\\ 
G &Amorphous carbon & 1.81 & $33.8$ & $78.7$ & - & 6.5 &  0.077 & 7.66 & Power-law GTD, $T_{\rm low}$ fixed.\\ 
H &Amorphous carbon & 1.81 & $19.8$ & $74.0$ & - & 4.1 &  0.22 & 7.79 & Power-law GTD, $T_{\rm low}$ free.\\ 
I &Amorphous carbon & 1.81 & $33.8$ & $83.9$ & 7.79 & - &  0.072 & 7.75 & Exponential GTD, $T_{\rm low}$ fixed.\\
J &Amorphous carbon & 1.81 & ${\it 2.73}$ & $76.2$ & 11.0 & - &  0.71 & 7.73 & Exponential GTD, $T_{\rm low}$ free.\\ [1mm]

    \hline
  \end{tabular}
  \end{center}
  \end{table*}

\subsection{Test case: fitting the Crab Nebula SED}
\label{testapp}
We have chosen to use the Crab Nebula to test our simple multi-temperature component model since it has a wide FIR/sub-mm SED which seems to suggest a range of grain temperatures is possible \citep[see also the papers by TD13 and][where a range of grain temperatures is discussed as well]{Richardson13}. SN 1987A, on the other hand, has a very narrow SED which appears consistent with a single-temperature population with an extremely low dust temperature \citep{Matsuura11} and thus makes it irrelevant as a test case in the present context. 

There is also a third well-observed remnant that has received a lot attention because of its seemingly large dust mass, Cas A \citep[see][and references therein]{Dunne09}, but Cas A is known to be a complicated case. It is far from established that the FIR/sub-mm part of the SED primarily reflects dust in the SNR, because of foreground contamination from a spiral arm \citep{Dunne03,Krause04,Gomez09,Barlow10}. Thus, we decided to {\it not} consider Cas A as a test case either, although the shape of the SED would suggest a fairly wide temperature distribution as seen in the Crab Nebula.

\begin{figure}
  \resizebox{\hsize}{!}{
   \includegraphics{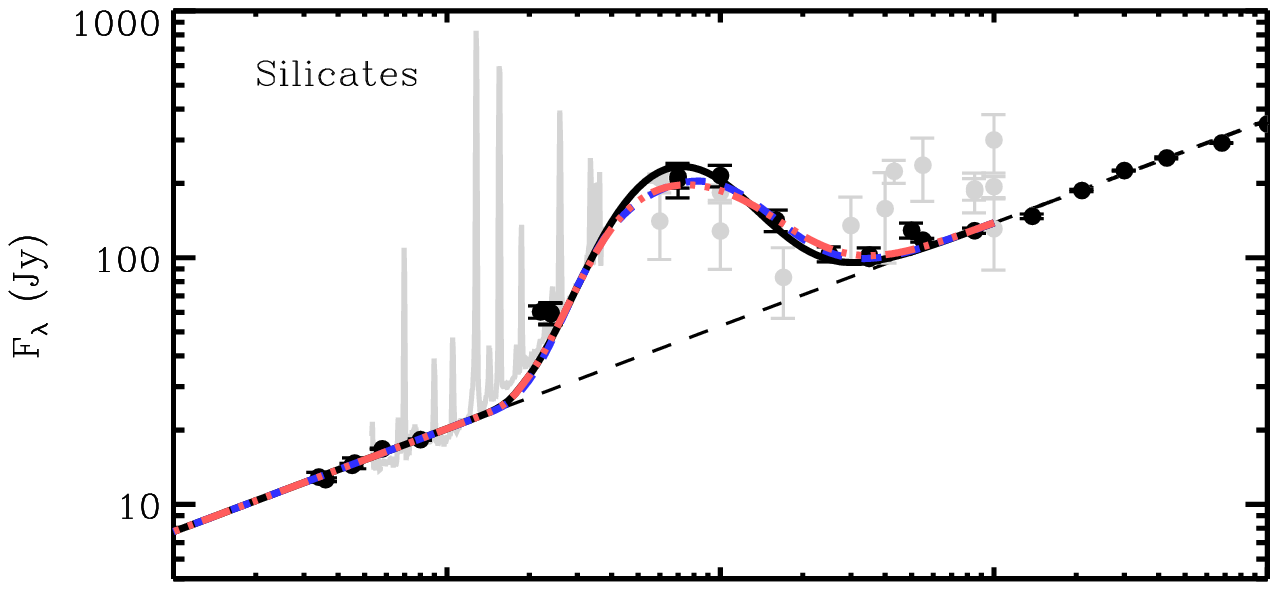}}
   \resizebox{\hsize}{!}{
   \includegraphics{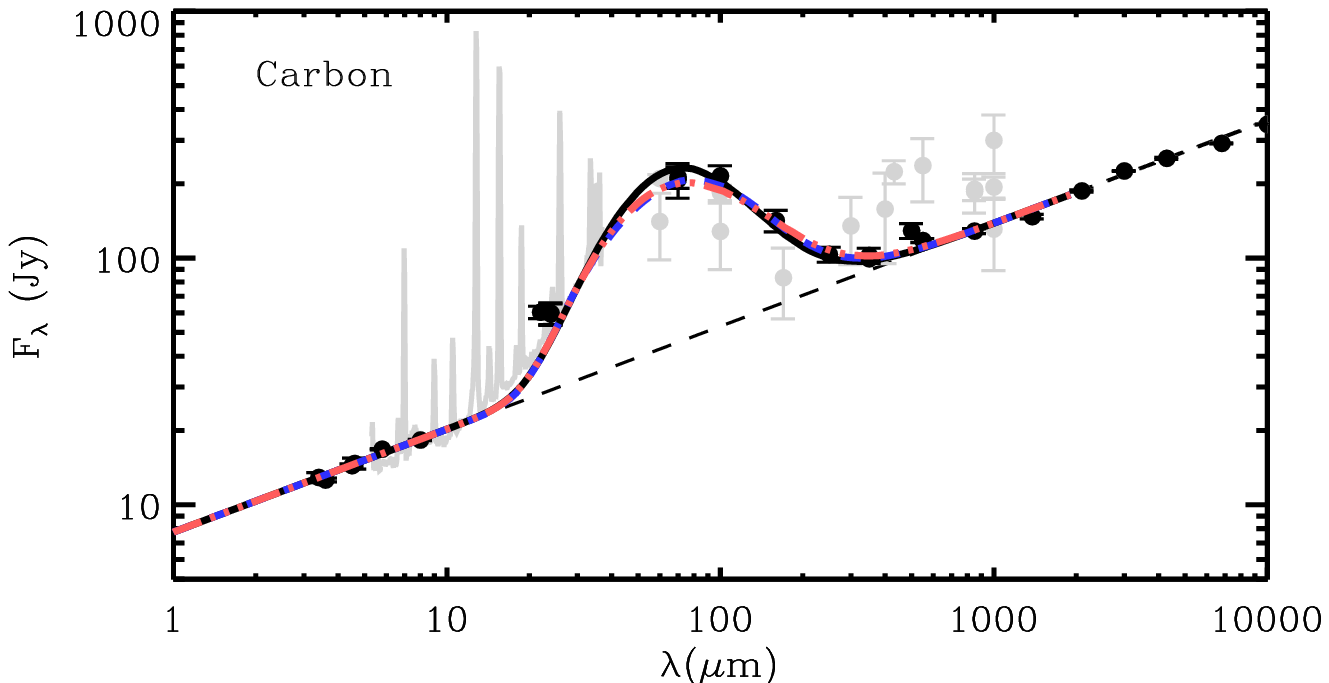}}
  \caption{\label{crabneb} { SED model fits to the Crab Nebula using a power-law GTD with photometric data from G12.  The dashed blue lines correspond to the two-component models by G12, while the solid black and read dot-dashed lines show temperature-distribution models with $T_{\rm low}$ fixed and as a firing parameter, respectively. The thin dashed black line shows the estimated synchrotron component. The upper panel show models based on silicate dust, while the lower panel shows models based on amorphous carbon dust.}}
  \end{figure}

We use the photometric data from G12 to plot the SED of the Crab Nebula (Figure~\ref{crabneb}) with wavelengths ranging from $1 -1000\,\mu$m. The Crab Nebula is a somewhat special object in that it has a strong synchrotron radiation field originating from the pulsar wind nebula (PWN). The PWN is the main heating source, a fact that TD13 take advantage of in their model (i.e., in this particular case, the heating source can also be specified quantitatively, but this is not the case for other SNRs). For our purposes, we only have to subtract the synchrotron component from the SED, which is relatively easy since it is well-described by a power-law (see G12 and references therein).  The integrated fluxes need also to be corrected for line emission. In some cases it is small, e.g., only 8.7\% of the total flux at $100\,\mu$m and 4.9\% of the $70\,\mu $m (see Table 2 in G12). But the $24\,\mu$m flux due to dust emission is 43\% of the synchrotron subtracted flux \citep{Temim12}, which is important to take into account.

\begin{figure}
  \resizebox{\hsize}{!}{
   \includegraphics{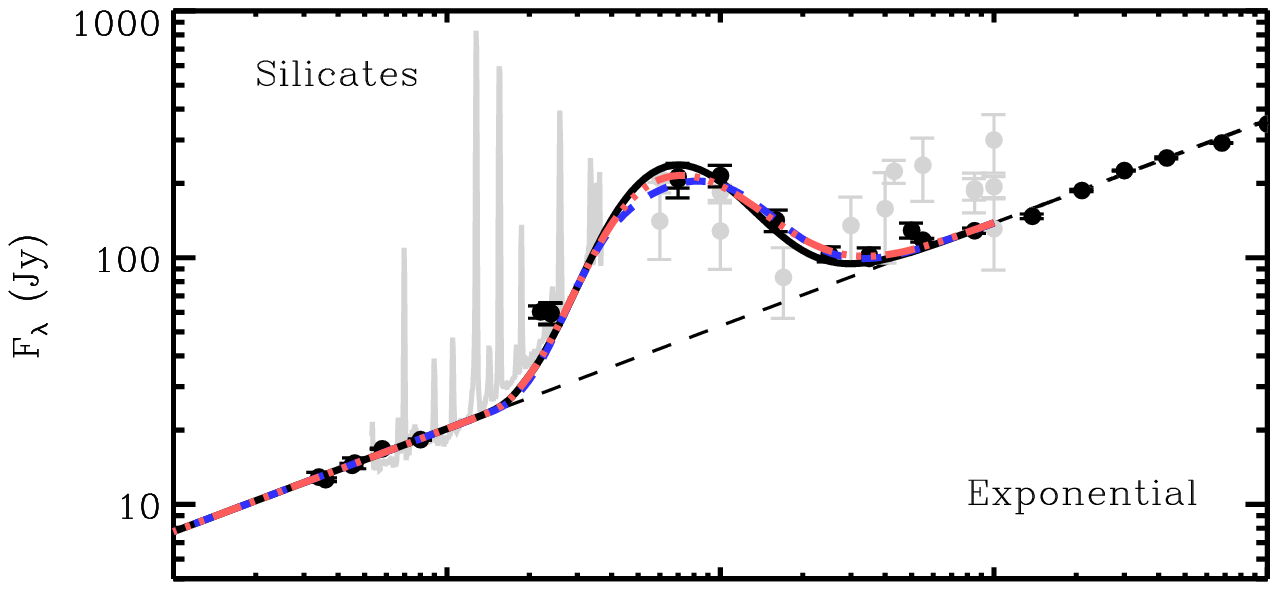}}
   \resizebox{\hsize}{!}{
   \includegraphics{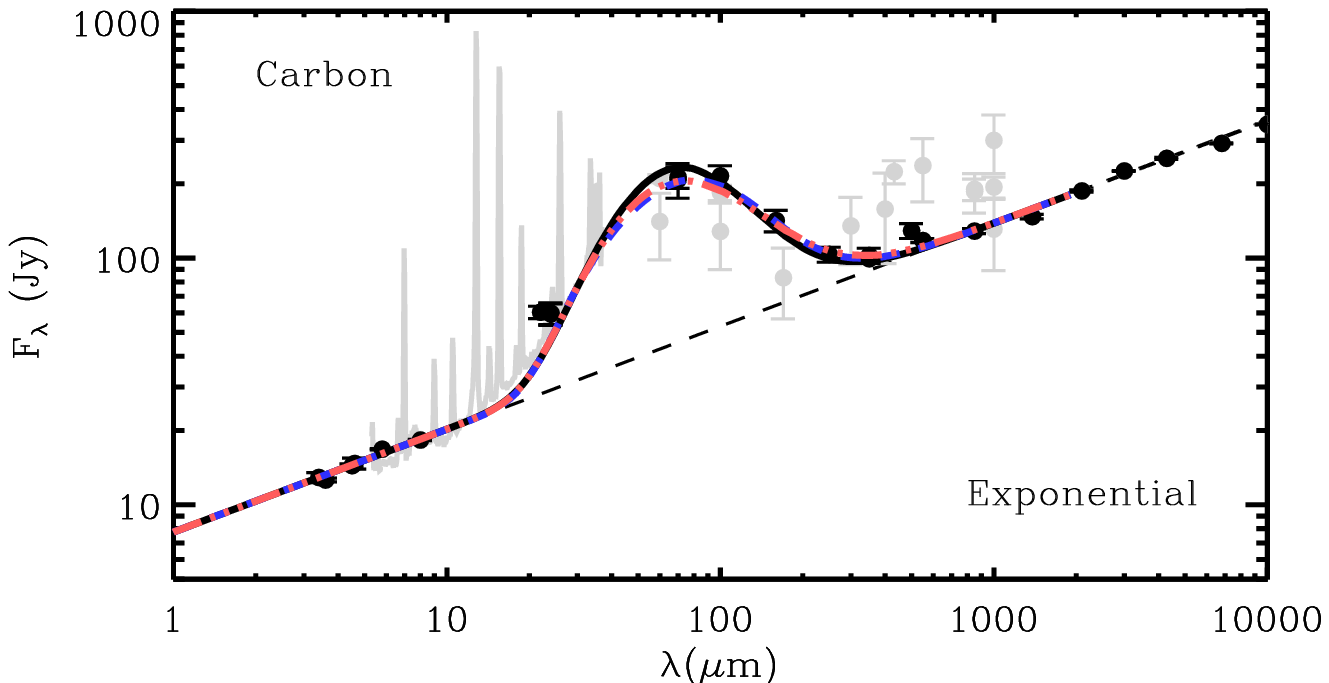}}
  \caption{\label{crabneb2} Same as Fig. \ref{crabneb} but for an exponential GTD.}
  \end{figure}

Since G12 obtained a valid fit with a canonical two-temperature fit with astronomical silicates \citep{Draine84,Weingartner01} and amorphous carbon \citep{Zubko96}, we first modelled the SED using astronomical silicates but now with the two forms of GTDs described in Section \ref{multitemp}. The range of dust temperatures (Table~\ref{parameters}) obtained through the SED fitting corresponds to a steep distribution favouring cold dust (see Fig. \ref{graintemp}). { Since the low-temperature limit $T_{\rm low}$ is decisive for the inferred mass, we have considered two cases: (1) a lower limit on the dust temperature set to the cold dust component temperature derived in G12 (again, see Table \ref{parameters}) and (2) a lower limit which is treated as free parameter, albeit with an absolute lower limit set by the cosmic background temperature $T_{\rm CMBR} = 2.73$~K. We will later discuss the effects of assumptions about $T_{\rm low}$ in more detail (Section 3.4).} The upper limit is treated as a free parameter for all cases, but is unsurprisingly rather similar to the temperature of the warm component as obtained from a two-temperature fit (see Table \ref{parameters}).  The models with a range of temperatures (full-drawn black and dot-dashed red curves in Figs. \ref{crabneb} and \ref{crabneb2}) are as good fits to the data (lower $\chi^2$) as the two-temperature fits (the blue dashed lines). But the corresponding dust masses are quite different. Assuming silicate dust and $T_{\rm low}$ fixed at the value obtained from the two temperature fits, only $\sim$50\% of the dust {\it mass} is required (compared to the two-temperature fit). Using amorphous carbon grains \citep[data taken from][]{Zubko96} with a range of temperatures instead of silicates yields similar results, though with a revised dust mass of 60-70\% compared to the carbon grain model in G12. The second case, where $T_{\rm low}$ is a fitting parameter, leads to lower $T_{\rm low}$ values and thus significantly higher dust masses. With a power-law GTD the dust masses, assuming silicates as well as amorphous carbon, are roughly doubled compared to the two-temperature fits. Using an exponential GTD, the fitting algorithm pushes $T_{\rm low}$ to the minimum value $T_{\rm low} = T_{\rm CMBR} = 2.73$~K. Maximising the amount of very cold grains like this suggests dust masses which are a factor 4-6 higher than those obtained in the two-temperature fits. Obviously, this last result is not very realistic, but it clearly demonstrates why fitting simple SED models to data can be dangerous.
 
The power-index values we obtain for the power-law models agree well with the values obtained by, e.g., \citet{Kovacs10,Magnelli12} who used a similar GTD approach for deriving the dust mass in galaxies.  \citet[][see also references therein]{Kovacs10} discussed that a power index in the range $\alpha = 6.5...8.5$ is expected in diffuse media, while for dense interstellar media $\alpha = 5...7$ is more likely. The upper end of these ranges correspond to an effective emissivity index $\beta = 2$, which is appropriate for, e.g., astronomical silicates. Thus, a SNR (which can be regarded as a dense medium) with silicate dust should have $\alpha = 7$, which is exactly the value we have obtained for the Crab Nebula with $T_{\rm low}$ fixed (see Table \ref{parameters}). For amorphous carbon dust, which has $\beta \approx 1$, we should expect $\alpha \approx 6$, in agreement with our results in Table \ref{parameters}. 

In principle, the dust masses obtained with a fixed $T_{\rm low}$ could lower the tension between the expected metal budget and the amount of dust formed in the ejecta. The mass of the progenitor star is constrained by the fact that one can put relatively strict constraints on the mass of the neutron star, $M_{\rm ns} \approx 1.4M_{\sun}$, which suggest a progenitor mass below $13\,M_{\sun}$ according to theoretical models \citep[see, e.g.,][]{WW95}. But we note also that the chemical abundances seem to be in better agreement with a progenitor above $11\,M_{\sun}$ \citep{MacAlpine08}. In general, it seems a progenitor of relatively low mass is favoured, which is supported by the slow expansion velocity \citep{Fesen97,Smith03}. A recent assessment by \citet{Smith13} has strongly constrained the progenitor mass to 8-10~$M_{\sun}$, suggesting a super-AGB star that suffered an electron-capture SN rather than an Fe core collapse event, which puts a severe limit on the metal budget. The masses found by G12 are then only marginally consistent with the metal budget if one accepts the nucleosynthetic models by \citet{WW95}. A $13\,M_{\sun}$ star in such case allows $0.37\,M_{\sun}$ of silicates (assuming an effective mass number $A_{\rm sil } = 170$) and $0.11\,M_{\sun}$ carbon dust to be formed. Thus, while the dust masses found by G12 are not obviously overestimates, a model suggesting close to 100\% dust-condensation efficiency is not very convincing since one expects a non-negligible sublimation rate in a SNR. The lower dust masses may seem like a reasonable and conservative choice, but one has to be aware that the total SED cannot provide conclusive evidence.

\subsection{How much difference does a multi-temperature fit make?}
\subsubsection{Shape of the SED}
The range of dust temperatures is directly connected to the width of FIR/sub-mm (dust) bump in the SED. `Warmer' grains (with temperatures above the coldest grains, that is) will inevitably add flux on the short-wavelength side of the dust SED. What this extra flux will look like in the SED depends on the dust type, though. Astronomical silicates can provide a tell-tale signature -- the $10\,\mu$m feature -- that reveals the presence of warm dust ($T_{\rm d}\gtrsim 100$\,K), while the carbon-dust FIR/sub-mm SED is mostly featureless. The $10\,\mu$m feature cannot arise from cold dust, however. But the location of the peak wavelength and slope of the short-wavelength tail poses a constraint on the cold-dust GTD regardless of the presence of any dust-emission features. 

The kurtosis (`peakedness') of the dust SED can also be affected by the GTD. In particular, using a continuous GTD may result in more flux in the middle of the SED than in the case of a two-temperature model. This is clearly seen in the fits to the Crab Nebula with a fixed low-temperature limit (see Section \ref{testapp}, Figs. \ref{crabneb} and \ref{crabneb2}). Thus, a continuous GTD may differ slightly from the two-temperature model in terms of the shape of the resultant SED. This affects the required dust mass of the model because the fit to the observed SED will be different. An increased kurtosis in the model SED should result in a lower dust mass, which is also what we obtained in our application to the Crab Nebula. 
%However, this effect depends on the observational uncertainty of the SED, i.e., how much `freedom' there is in the SED fit. 

    \begin{table}
  \begin{center}
  \caption{\label{grid} Parameter ranges for the grid of artificial SEDs.}
  \begin{tabular}{lll}
  \hline
  \hline
  Parameter & Range & Step size \\
  \hline
  $T_{\rm low}$   & 20 - 50\,K & 0.3\,K \\
  $T_{\rm high}$ & 60 - 100\.K & 0.6\,K \\
  $T_{\rm 0}$       & 3 - 9\,K  & 0.06\,K\\
  $\alpha$            & 4 - 9      & 0.05 \\
    \hline
  \end{tabular}
  \end{center}
  \end{table}
  
     \begin{figure*}
   \includegraphics[scale=0.29]{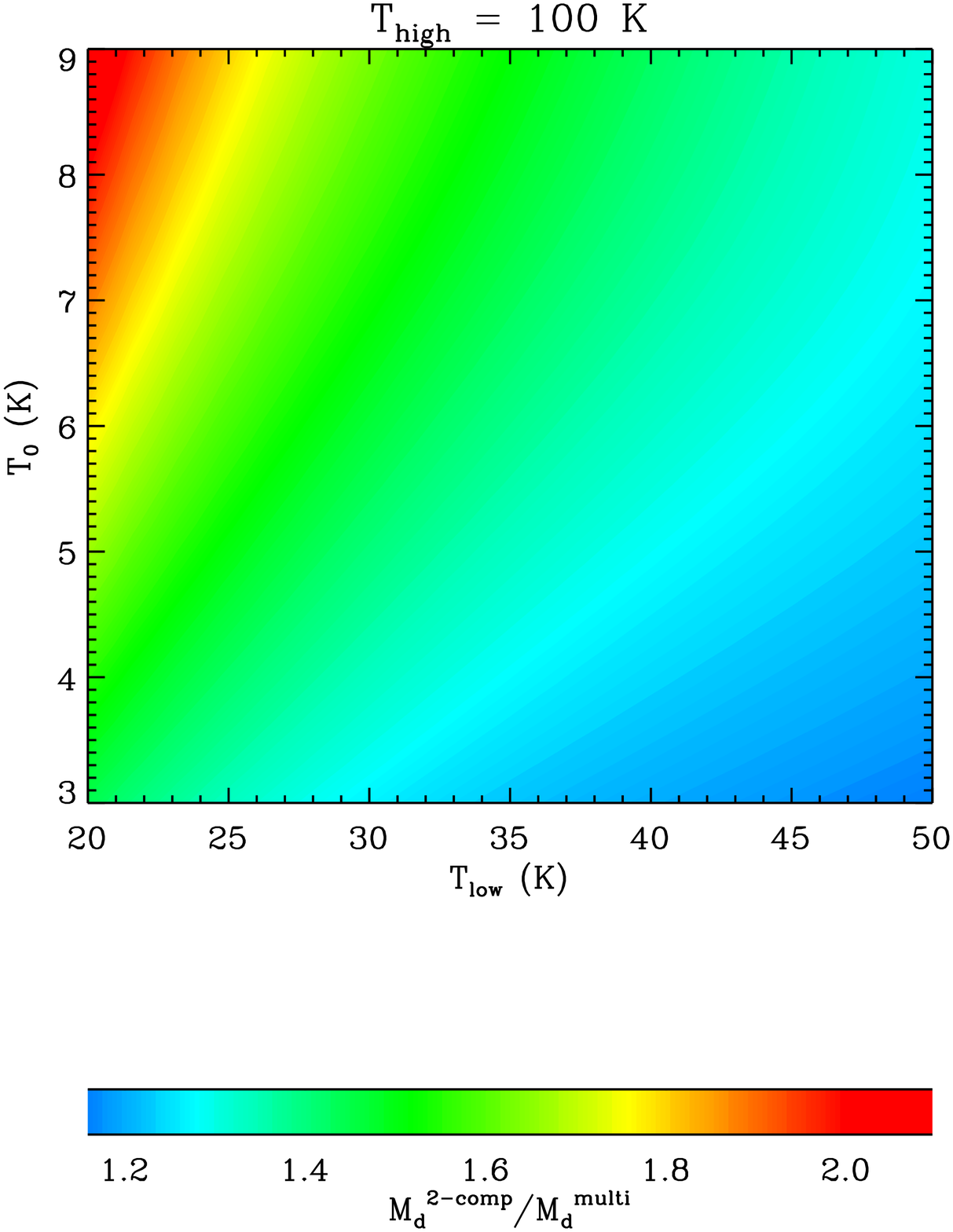}
   \includegraphics[scale=0.29]{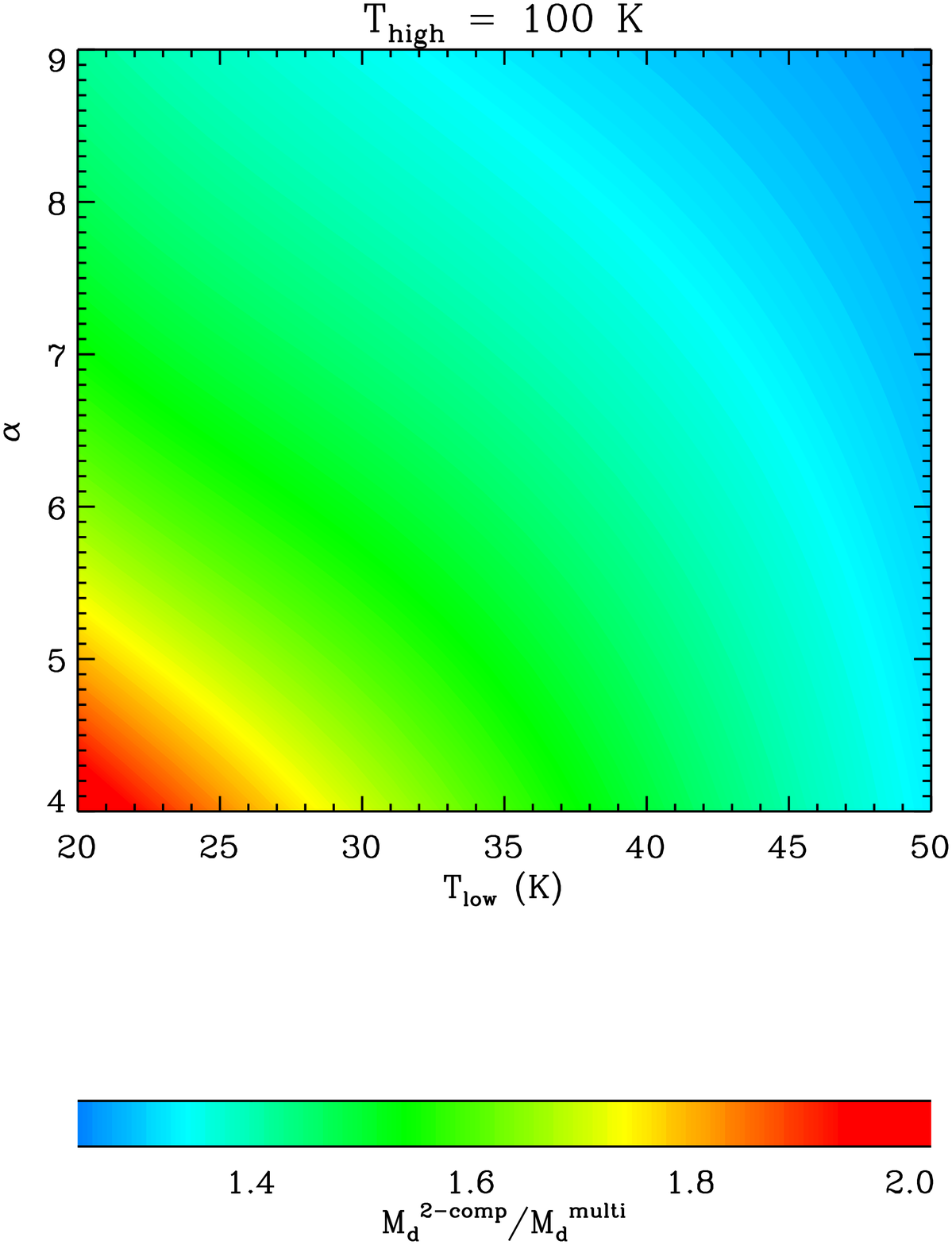}
  \caption{\label{massratio1} The implied dust masses from a two-temperature fit to a grid of SEDs generated from simple GTDs compared to the `true' dust masses corresponding to the adopted GTDs.
  Left panel: the effect on the dust mass assuming an exponential GTD with $T_{\rm high} = 100$\,K and $T_0$ the scaling temperature of the GTD. Right panel: same as left panel, but for a power-law GTD with power index $\alpha$. The temperature of the cold component in the two-temperature fits are assumed to be the same as the low-temperature cut-off in the GTD, which represents a reasonable upper limit to the grain size. 
  These figures may not display well in b/w, but are available in colour in the online version of the journal.}
  \end{figure*}
  
    \begin{figure*}
   \includegraphics[scale=0.29]{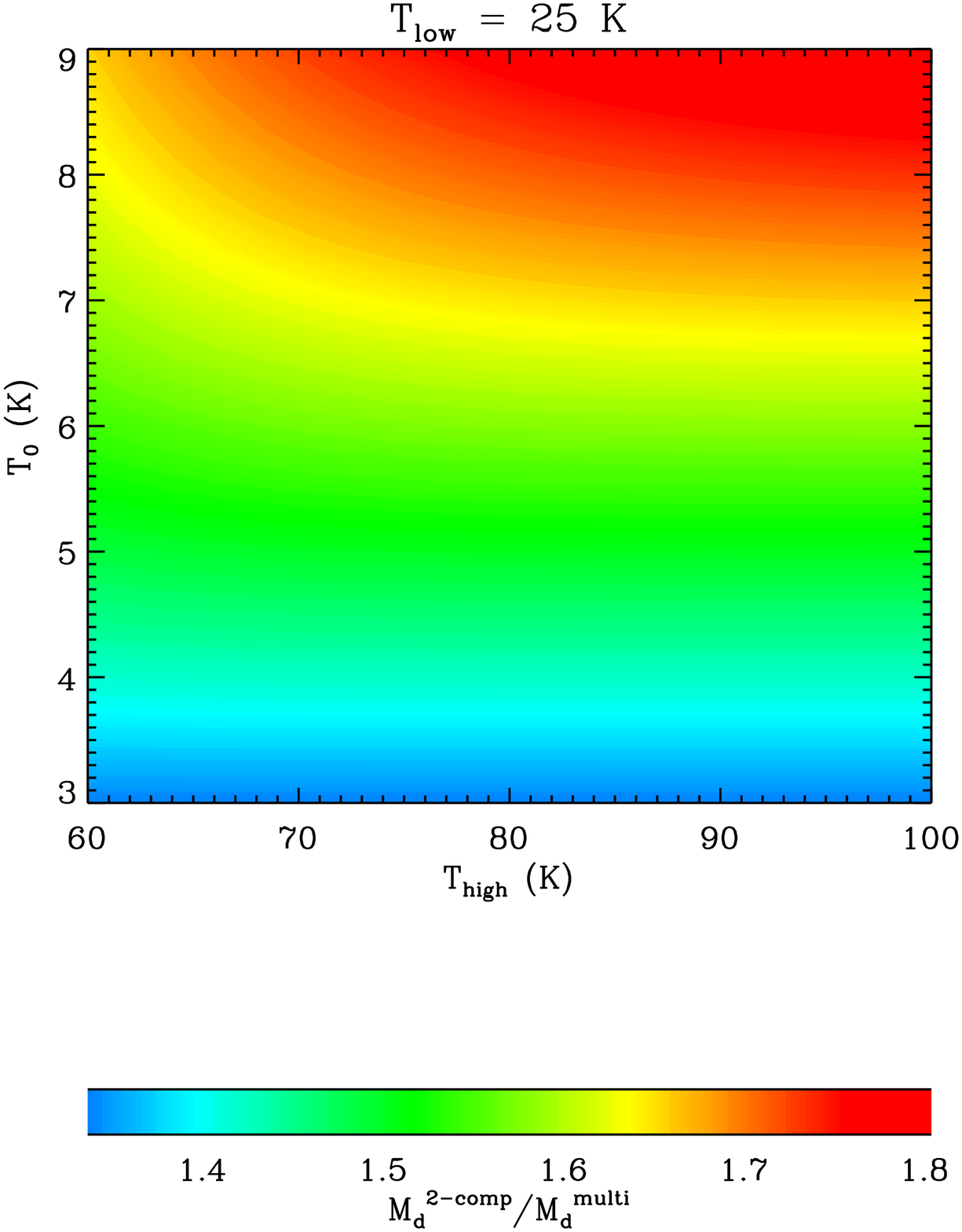}
   \includegraphics[scale=0.29]{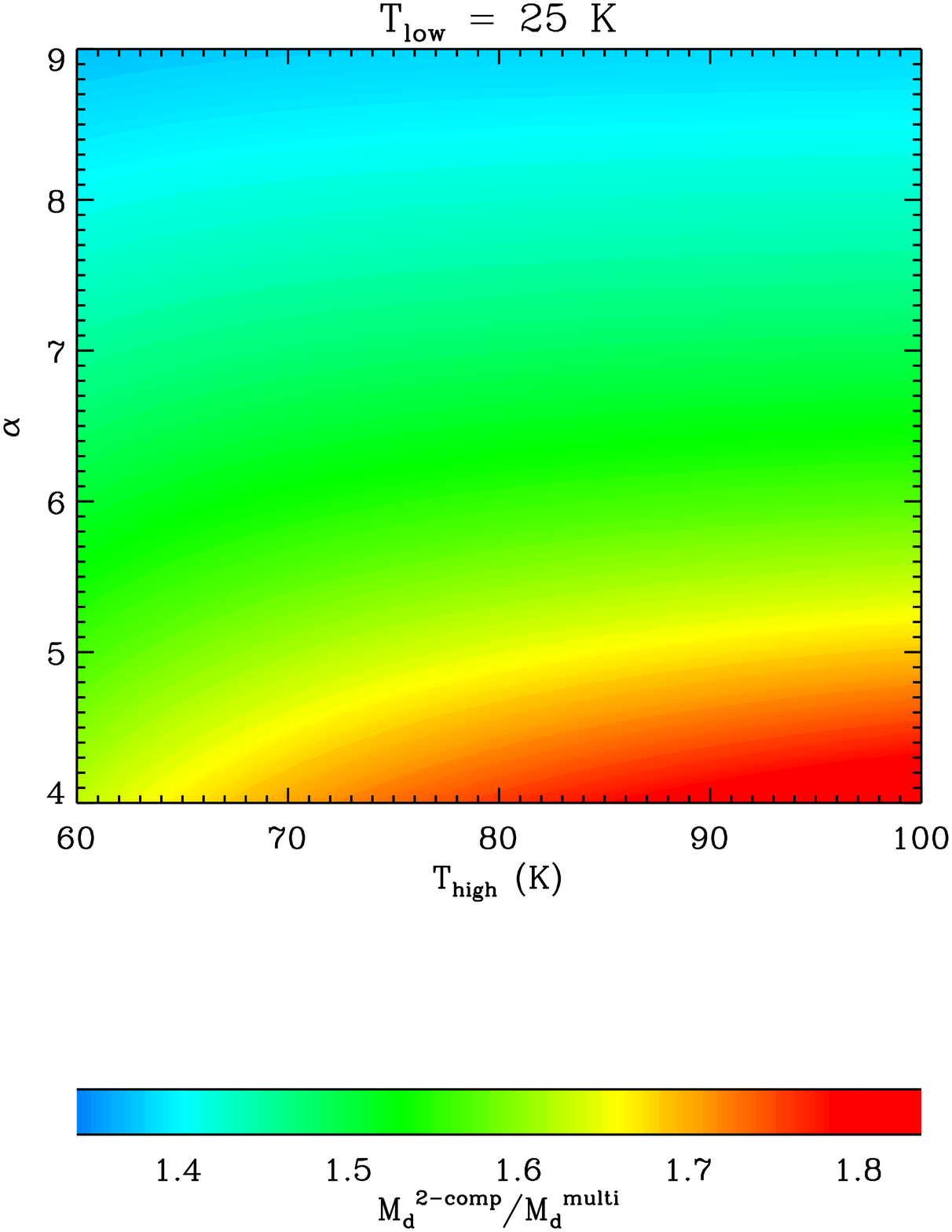}
  \caption{\label{massratio2} Same as Fig. \ref{massratio1} but with the lower temperature limit fixed at $T_{\rm high} = 25$\,K and variation of the upper limit. Left panel shows the case of an exponential GTD and the right panel a power-law.}
  \end{figure*}
  
     \begin{figure*}
   \includegraphics[scale=0.29]{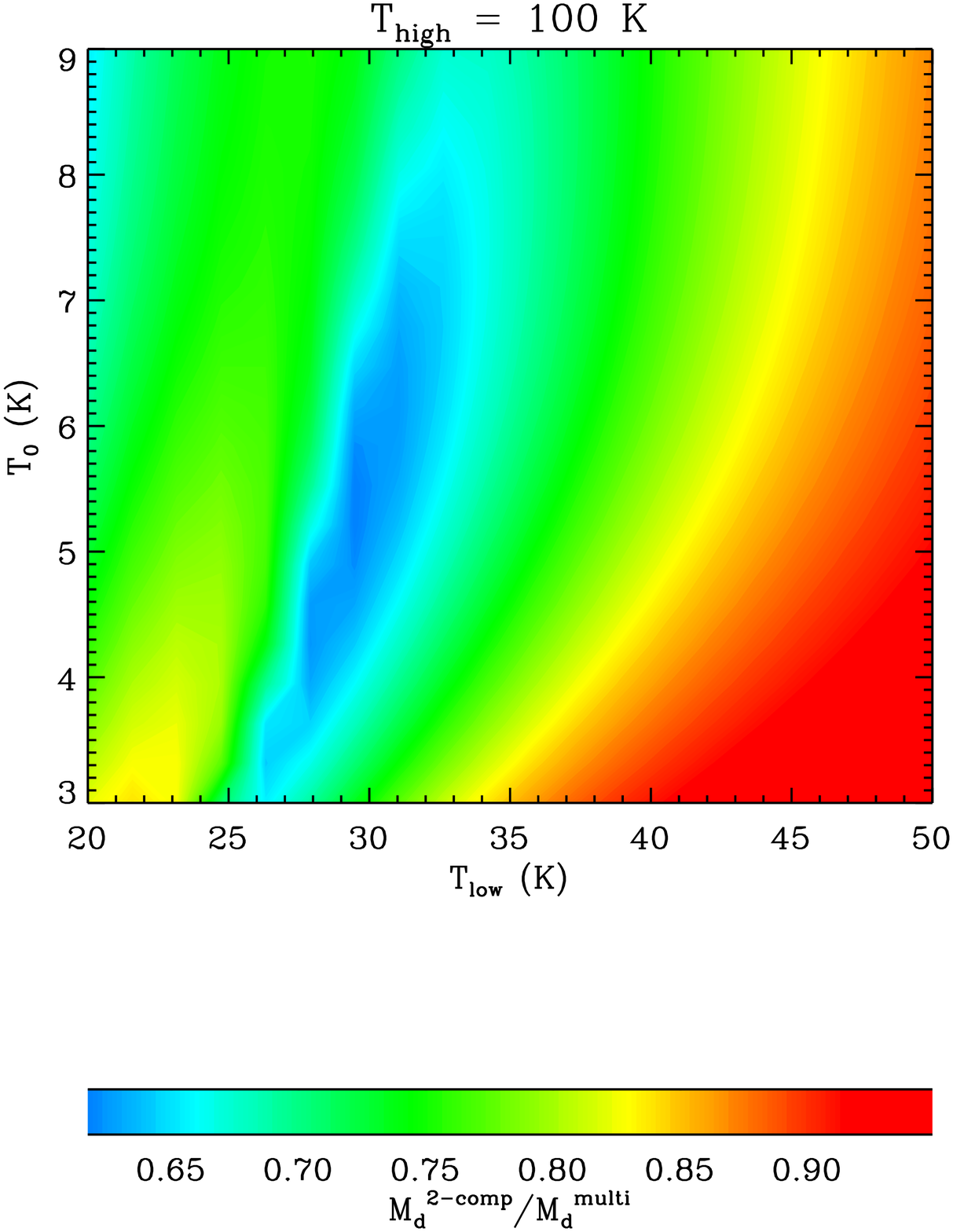}
   \includegraphics[scale=0.29]{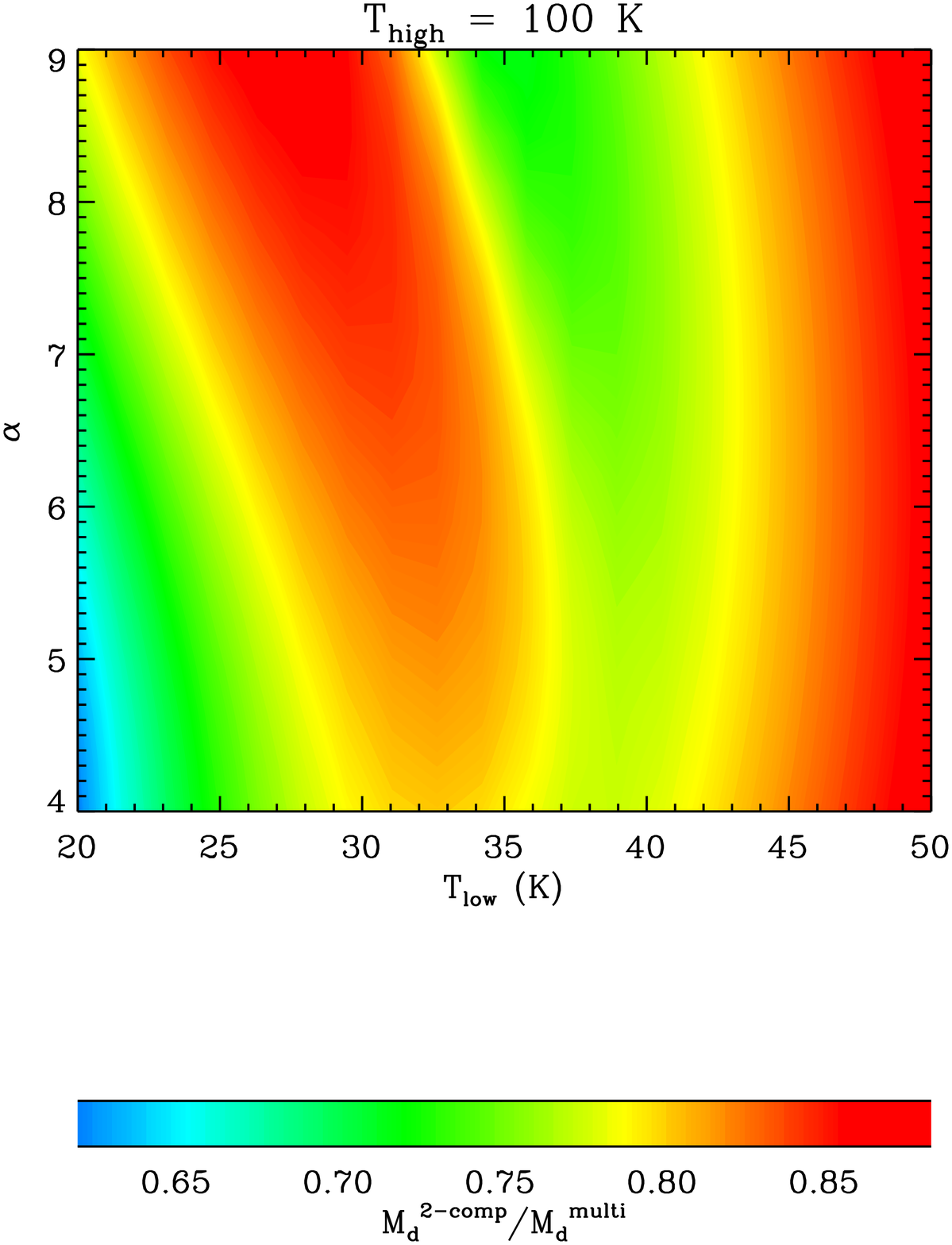}
  \caption{\label{massratio3} Same as Fig. \ref{massratio1} but without assuming that the temperature of the cold component in the two-temperature fits is the same as the low-temperature cut-off in the GTD.}
  \end{figure*}
  
    \begin{figure*}
   \includegraphics[scale=0.29]{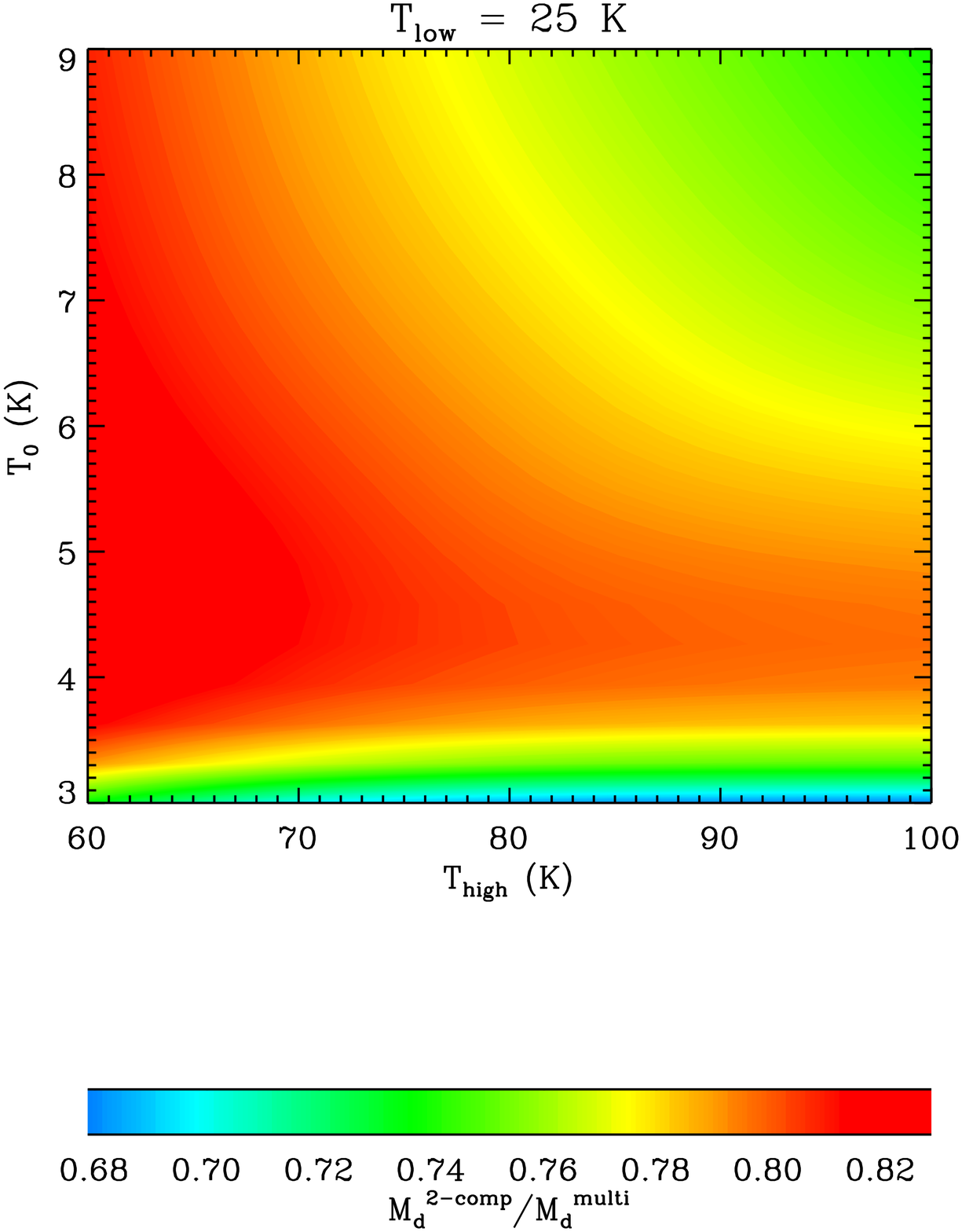}
   \includegraphics[scale=0.29]{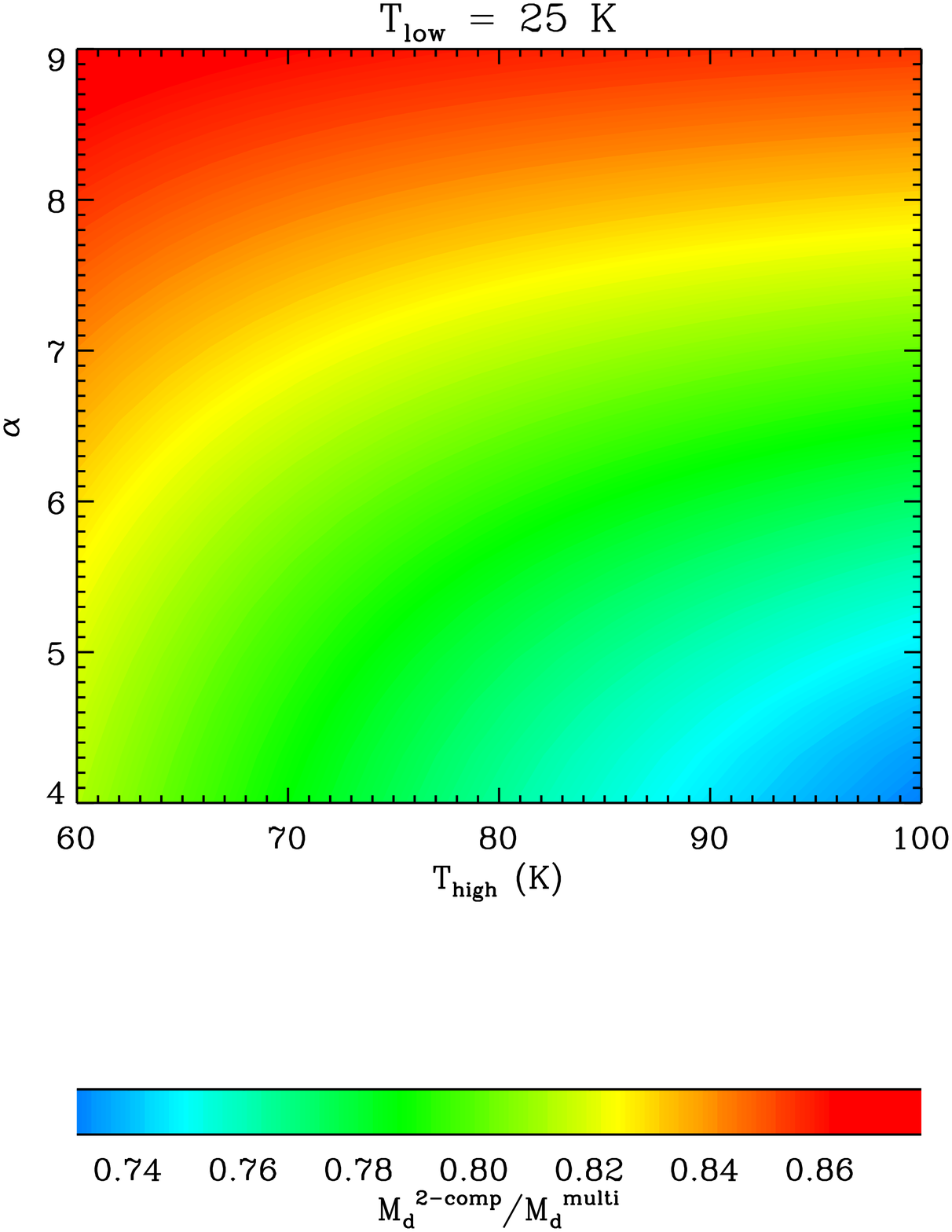}
  \caption{\label{massratio4} Same as Fig. \ref{massratio1} but without assuming that the temperature of the cold component in the two-temperature fits is the same as the low-temperature cut-off in the GTD.}
  \end{figure*}

\subsubsection{Reducing the dust mass?}
\label{reducing}
In the continuous case (the number of temperature components $N\to\infty$), we can write $S_\lambda$ as a temperature mean of the Planck function $B_\lambda$ weighted by the GTD $W(T_{\rm d})$,
\begin{equation}
\label{slambda}
S_\lambda = \left[ \int_{T_{\rm low}}^{T_{\rm high}} W(T_{\rm d})\,dT_{\rm d}\,\right]^{-1} \int_{T_{\rm low}}^{T_{\rm high}} W(T_{\rm d})\,B_\lambda(T_{\rm d})\,dT_{\rm d}, 
\end{equation} 
which replaces $B_\lambda$ in the derivation of the dust mass from a given SED. The total dust mass obtained from a two-temperature fit is completely dominated by the cold component (see, e.g., G12) and, as we have pointed out above, it seems reasonable to associate the low temperature cut-off $T_{\rm low}$ with the temperature of the cold component $T_{\rm cold}$ from the two-temperature fit (although this is not obviously the case, which we will discuss later). Hence, if we compare $S_\lambda(T_{\rm low},T_{\rm high};T_0)$ and $B_\lambda(T_{\rm low})$, we would have an approximate measure of how much the GTD affects the implied dust mass for a given set of parameters, if the cold-dust  temperature is equal to the lower temperature limit  $T_{\rm low}$ of the GTD. Since we are considering thermal radiation at long wavelengths, we may also approximate the Planck function using the fact that $e^{1/x}-1\approx x$ for $x\gg 1$. It is then straight forward to show that, for $W(T_{\rm d}) = W_0\exp(-T_{\rm d}/T_0)$, $T_0 \ll T_{\rm high}$ (very steep GTD)
%, we have
%\begin{equation}
%\label{sbratio}
%{S_\lambda\over B_\lambda} \approx  {(1+ T_{\rm low}/T_0)\,e^{-T_{\rm low}/T_0}-(1+ T_{\rm high}/T_0)\,e^{-T_{\rm high}/T_0}\over T_{\rm low}/T_0\,(e^{-T_{\rm low}/T_0} - e^{-T_{\rm high}/T_0})}.
%\end{equation}
%We note that in the case $T_0 \ll T_{\rm high}$ (a very extended range of grain temperatures and/or a steep GTD) 
we have at large wavelengths
\begin{equation}
\label{sbratio2}
{S_\lambda\over B_\lambda} \approx 1+ {T_0\over T_{\rm low}}, 
\end{equation}
which is because the effect of a wide range of dust temperatures is in such case limited and depends on the balance between cold and warm dust. That is, the effect of a realistic GTD (which must be steep) on the implied dust mass depends mostly on the temperature of the coldest dust. Moreover, using Eq. (\ref{sbratio2}) and since $T_{\rm low}$ is typically a few times $T_0$ (if the temperature $T_{\rm low}$ is that of the two-temperature fit), one can easily verify that the increase in flux from a GTD compared to the two-temperature fit is not likely more than $20-30$\%. 

We have also computed Eq. (\ref{slambda}) by numerical integration using an exponential as well as an power-law GTD with $Q_{\rm abs}$ for `astronomical silicates' \citep{Weingartner01} and amorphous carbon \citep{Zubko96}. Because this integration is computationally fairly inexpensive, we generated a large, dense grid of artificial SEDs with either a fixed lower temperature limit or a fixed upper temperature limit and a range of values for the remaining parameters (see Table \ref{grid}).  For each SED we then fitted two-temperature models with the temperature of the cold component fixed to $T_{\rm cold} = T_{\rm low}$ and with $T_{\rm cold}$ as a free parameter. We then compared the required dust masses, i.e., we computed the ratio $M_{\rm d}^{\rm 2-temp}/M_{\rm d}^{\rm multi}$.  These ratios are shown in Figs. \ref{massratio1}, \ref{massratio2},  \ref{massratio3} and \ref{massratio4}.   In total we computed $4\cdot 10^{6}$ SEDs and made two-temperature fits all of these, but we only show a subset of 80000 SEDs here (since the results for silicates are very similar to those for amorphous carbon and we do not display all the various cases of fixed upper and lower temperature limits either).

As can be seen on Figs. \ref{massratio1} and \ref{massratio2}, the $M_{\rm d}^{\rm 2-temp}/M_{\rm d}^{\rm multi}$-ratios suggests the introduction of a GTD (with $T_{\rm low}=T_{\rm cold}$) only lowers the implied dust mass by $\sim 30-40$\% for realistic values of $T_{\rm low}$, $T_{\rm high}$ and $T_0$ in  case of an exponential GTD (left panels in Fig. \ref{massratio1} and \ref{massratio2}) and a similar result is obtained for the power-law case (right panels in Fig. \ref{massratio1} and \ref{massratio2}). Comparing the Figures, it is also clear that the lower temperature limit $T_{\rm low}$ is more important than the upper limit $T_{\rm low}$ for any reasonable GTD (which must have $T_0 \ll T_{\rm high}$ or $\alpha \sim 5...8$), as we predicted from Eq. (\ref{sbratio2}) above. 

The $\sim 50$\%  (or $30-40$\%) reduction of the (silicate or carbon) dust mass that we obtained by including a GTD in the SED modelling of the Crab Nebula is in fact due to two factors that make comparable contributions: the increased flux due to the addition of a range of grain temperatures (higher average $T_{\rm d}$) and a mere difference in the fit due to the slightly different shape of the model SED. The latter can be seen in Figs. \ref{crabneb} and \ref{crabneb2}, as well as Fig. 5 in TD13. The combined effect amounts to a {\it factor of two} for silicates in the Crab Nebula, which is likely the largest reduction one can expect in employing a GTD in general (not only for the Crab Nebula).  This is expected since in case of an exact fit to the SED, the relative mass reduction would be the same as the relative flux increase from adding more warmer dust (typically $20-30$\%). As is evident from figure \ref{massratio1} and \ref{massratio2}, the expected dust-mass ratio, comparing a two-temperature fit with a GTD-model fit, would only in extreme cases (e.g., very cold dust) reach above two.
  
It is worth stressing that the mass reduction we describe above is {\it totally dependent} on the assumption that the temperature of the cold-dust component in a two-temperature fit is the same as the lower temperature limit of the GTD ($T_{\rm low}=T_{\rm cold}$). This assumption is similar to assuming an upper-limit grain size, as in TD13, and has essentially the same effect: it limits the amount of dust mass. However, we performed further numerical experiments which showed that if one allows the GTD to reach down to very low dust temperatures, a two-temperature fit to that GTD may actually predict the same or even a {\it higher} dust mass, compared to the two-temperature fit by G12. The effect of the extra flux from warmer dust on the predicted dust mass may, in reality, be counteracted by the presence of dust colder than the cold component of a two-temperature fit. Such cold grains may affect the dust mass without making a significant contribution to the SED, thus still yielding a good model fit. 

In Figs. \ref{massratio3} and \ref{massratio4} we show the implied dust masses from a two-temperature fit to a grid of SEDs generated from simple GTDs compared to the `true' dust masses corresponding to the adopted GTDs as in Figs. \ref{massratio1} and \ref{massratio2}, except that the temperature of the cold component in the two-temperature fits is here treated as a {\it free} fitting parameter\footnote{Note that this is different from the Crab Nebula fits with $T_{\rm low}$ as a free parameter, although it illustrates the same phenomenon.}. The dust masses inferred from the two-temperature fits are typically $\sim 10-40$\% lower than the `true' dust masses. The cold component temperature is typically $5-10$\,K higher than the low-temperature cut-off in the GTD, although there is no general scaling relation (see the contours in Figs. \ref{massratio3} and \ref{massratio4}). With a fixed high-temperature cut-off at $T_{\rm high} = 100$\,K the low-temperature range $T_{\rm low} = 25 - 35$\,K represents a special case: for an exponential GTD the two-temperature dust mass has a local minimum, and for a power-law GTD there is local maximum in this $T_{\rm low}$ interval. This special low-temperature range depends somewhat on the choice of $T_{\rm high}$ and for narrow GTDs such a temperature range may not exist. But it coincides with the range of dust temperatures that is the most interesting for cold dust in supernovae. It seems, therefore, that the effect on the inferred dust mass depends on the form of the GTD, which indicate that detailed radiative transfer models may be the best option as it will provide the best possible information about the GTD. Overall, our numerical experiments presented here, suggest that a two-temperature fit may not be as bad as it may seem at first, judging from the results of TD13 as well as our results presented in Figs. \ref{crabneb} and \ref{crabneb2}, but simple SED fits are still not reliable.

\subsection{Temperature-size relation}
\label{tdarel}
In an equilibrium model, there is a direct relation between grain size and grain temperature for a given (invariant) radiation field, which should be evident from Section \ref{radheat} (but see also Appendix \ref{hecoeq}). In a simple model, like the one we have used here, it is not meaningful to discuss grain sizes in quantitative terms based on the steepness of the GTD and temperature cuts as obtained from SED fitting. The reason is that Eq. (\ref{ebalance1}) is a local relation and adopted GTD is a global approximation. But locally the grain size must uniquely determine the grain temperature as long as nothing breaks the equilibrium. The steepness of the GTD is also directly dependent of $dT_{\rm d}/da$, i.e., the temperature-size relation is fundamental to the connection between the dust mass and the dust SED (see Eq. \ref{powerlaw2} and Appendix \ref{hecoeq}). But one should also remember that there is a degeneracy between the upper size limit of the grains and the slope of the GSD, which can become problematic even in models with detailed radiative transfer \citep[see][]{Owen15}. 

The upper temperature limit (corresponding to lower size limit) may have bearing on the implied dust mass obtained from the SED (see Figs. \ref{massratio2} and \ref{massratio4}) since the small grains usually only make up a small fraction of the dust mass, but still contribute significantly to the SED since they are slightly warmer than the large grains (the flux from a grain is essentially proportional to the temperature). TD13, on the other hand, argue that the choice of $a_{\rm min}$ has no particular effect on their results for the Crab Nebula (according to their Monte Carlo simulation) which is a result we can only reproduce when $T_{\rm low}$ is fixed and the GTD has a certain slope. Possibly, this is related to the fact that their model yields $dT_{\rm d}/da \approx 0$ for small grain radii $a$, though it is unclear why $T_{\rm d}$ becomes size independent for small grains in their model, which is also an equilibrium model. 
%Moreover, it is also unclear how $dT_{\rm d}/da \approx 0$ in the small-grain end is supposed to be interpreted from a physical point of view, since it would suggest a steep rise of the GTD at `high' grain temperatures, given a power-law GSD for all considered grain sizes.  

\section{Discussion and Conclusion}
Since there is known to be a distribution of dust-grain sizes, there must also be a distribution of grain temperatures (GTD) -- even in case of thermal equilibrium, which should apply to cold dust. We illustrated how this can increase the emission from any type of dust component, whether it is silicates, carbonaceous dust or another composition. We applied a GTD model to the SED of the Crab Nebula, which can be explained using thermal emission from solid dust grains at a range of dust temperatures (rather than the canonical two-temperature component model). A range of dust temperatures lower the required dust mass by $\sim$50\% and 30-40\% for astronomical silicates and amorphous carbon grains compared to recently published values ($0.25M_{\sun} \to 0.14M_{\sun}$ and $0.012M_{\sun}\to 0.0085M_{\sun}$, respectively), but the implied dust mass can also increase by as much as almost a factor of six ($0.25M_{\sun} \to 1.14M_{\sun}$ and $0.12M_{\sun}\to 0.71M_{\sun}$) depending on assumptions regarding the sizes/temperatures of the coldest grains. In general, we find/confirm that:
\begin{enumerate}
\item {The width of the SED determines how much small warm grains can contribute to the FIR/sub-mm flux excess. Only SNRs with relatively wide SEDs may show GTD effects worth considering.}\\[-2mm]
\item {Introducing a GTD has a significant but limited effect on the derived dust mass compared to a two-temperature fit. With the lower temperature limit of the GTD set equal to temperature of the cold component of the two-temperature fit, the implied dust mass is typically $\sim 50$\% larger for the two-temperature fit. But without this coupling there is not necessarily any dust-mass reduction resulting from the introduction of a GTD (the inferred dust mass may in fact {\it increase}).}\\[-2mm]
\item {The difference in shape of the SED (e.g., its `peakedness') for a two-temperature model compered to a GTD model, in combination with the flux uncertainties, lead to different fitting results, which may affect the implied dust mass as much as the extra flux from warmer grains added due to the GTD.}
\end{enumerate}

We have therefore shown that introducing a GTD may predict a different dust mass than a canonical two-component model. But we also demonstrate that the effect is limited: we have shown that this depends strongly on the temperature of the coldest dust and how well-constrained the observed SED is in the FIR/sub-mm. It is not appropriate to claim that introducing a GTD (or a grain-heating model as in TD13) gives us better constraints on the dust mass of a SNR, such as the Crab Nebula. First, we cannot know whether there should be a low-temperature limit in the GTD with a value similar to the temperature of the cold component in a two-temperature fit, even if this seems a reasonable assumption. The dust mass is uncertain by at most a factor of a few due to this. Second, the distances to many Galactic SNRs are relatively uncertain. The Crab Nebula is no exception: G12 adopt $D = 2$~kpc, which is exactly in the middle of the range $D = 1.5 - 2.5$~kpc given by \citet{Kaplan08}. Assuming $D = 2.0\pm0.5$~kpc the uncertainty of the dust mass is almost a factor of three. Third,  uncertainties in the optical and structural properties of the dust component amounts to at least a factor of a few: the emissivity of dust may vary considerably according to some observational estimates \citep[see, e.g.,][]{Alton00, Alton04,Dasyra05} and if volatiles (e.g. ice mantles, which we will discuss in a forthcoming paper) makes up a significant part of the dust, there may be an additional factor of two in the uncertainty.  Combining all of the above, the total (maximum) uncertainty range spans at least an order of magnitude, somewhat depending on available constraints.  A major point is that while the present GTD approach is not more accurate than the two-temperature approach, it demonstrates that simple SED fitting cannot really constrain the dust mass.

One may then ask whether the GTD is ultimately an improvement.  Here we would argue that incorporating a more physical model for dust heating and FIR/sub-mm emission from dust grains is indeed always a qualitative improvement, but the most concerning uncertainty in the model -- what the dust is actually made of -- still remains.  However, we have demonstrated that the GTD does not always reduce the dust mass derived from SEDs (as implied by TD13), but also that the GTD-related uncertainty is typically not a dominant source of uncertainty i.e. the effect of grain temperatures is usually a part of the total uncertainty in deriving dust masses from the SED.

The degree of dust condensation (the fraction of condensible material that end up in dust grains) is unlikely close to 100\%, but the uncertainties in the conversion from FIR/sub-mm flux to dust mass and the amount of metals available for dust formation prevent precise estimates. In conclusion, the efficiency of dust production in supernovae remains poorly constrained, even if we would construct a sophisticated model of dust emission. This does of course not mean that radiative transfer models based on dust-grain populations with a range of grain sizes and temperatures are not important (and needed) tools for converting infrared to submm SEDs to dust masses. We conclude that one should avoid using simple SED fits as far as possible whenever there is evidence of a range of grain temperatures. But the overall error in the amount of dust formed in supernova is still largely due to the uncertainties in dust composition, structure and behaviour of optical constants and thus determining the mass of dust remains extremely difficult.

\section*{Acknowledgments}
We thank Thomas Wilson and another, anonymous, reviewer for helpful suggestions and comments.
We also thank Darach Watson for helpful comments and stimulating discussions which have been essential for this work. 
The Dark Cosmology Centre is funded by the Danish National Research Foundation.
Nordita (Nordic Institute for Theoertical Physics) is funded by the Nordic Council of Ministers, the Swedish Research Council, and the two host universities, the Royal Institute of Technology (KTH) and Stockholm University.

\appendix

\section{Heating, cooling and equilibrium}
\label{hecoeq}
If the dominant source of heating is radiation (typically in the UV/optical), the dust grains may be regarded as being in local thermal equilibrium with the mean intensity of the radiation field around them, i.e., for grains of a specific radius $a$ and temperature $T_{\rm d}$ we have an energy gain (heating) due to absorption given by
\begin{equation}
\left({dE\over dt}\right)_{\rm abs} =  \int_0^\infty \pi\,a^2\,Q_{\rm abs}(\lambda,a)\,{J}_{\star,\lambda} \,d\lambda,
\end{equation}
and an energy release (cooling) due to emission given by
\begin{equation}
\left({dE\over dt}\right)_{\rm em} =  \int_0^\infty 4\pi\,a^2\,Q_{\rm abs}(\lambda,a) \,\pi B_{\lambda}(T_{\rm d}) \,d\lambda,
\end{equation}
where $J_{\star,\lambda}$ is the mean intensity of the local radiation field, $B_{\rm \lambda}$ is the Planck function. We could easily obtain $(dE/dt)_{\rm abs} = (dE/dt)_{\rm em}$ if the surrounding medium is optically thin in the wavelength range where the energy/radiation is absorbed/released and collisional heating is negligible. The optical depth of the medium affects the energy absorption but has little effect on the re-emission at long wavelengths, i.e., where emission can be described using the Rayleigh limit. Thus, in that limit and in thermal equilibrium we have
\begin{equation}
\label{ebalance}
 \int_0^\infty Q_{\rm abs}(\lambda,a)\,{J}_{\star, \lambda} \,d\lambda \approx 4\pi a \int_0^\infty Q'_{\rm abs}(\lambda) B_{\lambda}(T_{\rm d}) \,d\lambda,
\end{equation}
where $Q'_{\rm abs} = Q_{\rm abs}/a$. The emission at long wavelengths tends to follow a power-law in $\lambda$. Replacing $Q'_{\rm abs}$ with $Q'_0\,(\lambda/\lambda_0)^{-\beta}$ we may then write
\begin{equation}
\mathcal{J}_\star (a) \equiv \langle Q_{\rm abs} \rangle_{\star}\,{J}_{\star}  \approx 4\pi a \int_0^\infty Q'_0\left({\lambda\over\lambda_0} \right)^{-\beta} B_{\lambda}(T_{\rm d}) \,d\lambda,
\end{equation}
where we have defined the mean value
\begin{equation}
\langle Q_{\rm abs} \rangle_{\star} \equiv {1\over J_\star}\int_0^\infty Q_{\rm abs}(\lambda,a)\,{J}_{\star, \lambda} \,d\lambda, \quad J_{\star} \equiv \int_0^\infty {J}_{\star, \lambda} \,d\lambda.
\end{equation}
$J_\star$ is the wavelength-integrated mean intensity (regarded as constant in the present context) and $\beta \in [1,2]$ for all common types of dust. The constant $Q'_0$ can be determined by considering $Q'_{\rm abs}(\lambda_0)$ for a specific dust type 
which conforms to a power-law approximation with a constant $\beta$. Now, the integral over wavelength on the right-hand side can be evaluated analytically, which yields
\begin{equation}
\label{heating_A}
{\mathcal{J}_{\star}(a)\over a} \approx  {4\pi\, Q'_0 \lambda_0^{\beta}} \tilde{\sigma}_{\rm SB}\,T_{\rm d}^{4+\beta},
\end{equation}
where $\tilde{\sigma}_{\rm SB}$ is not the usual Stefan-Boltzmann's constant, but the corresponding number for a modified blackbody. In case $\beta = 0$ we would have $\tilde{\sigma}_{\rm SB}$ equal to the usual Stefan-Boltzmann constant. By Eq. (\ref{heating_A}) it is clear that grain temperature should be weakly (anti-)correlated with grain size, though this correlation depends slightly (but not much) on the heating source and the validity of the approximations used above of course \citep[cf. Table 3 in][]{Draine84}. Note that an equilibrium model such as the one above is strictly valid only locally. 

%We have now arrived at the following rather obvious result:\\[1mm]
%I. {\it For dust grains kept in thermal equilibrium by an invariant radiation field, the grain temperature is uniquely determined by the grain size.}\\[1mm]

If heating is due to short-wavelength radiation, we are close to the grey-absorption limit (particles are large compared to the wavelength) in which case $\langle Q_{\rm abs} \rangle_\star = 1$ and thus $\mathcal{J}_{\star} = J_\star = {\rm constant}$. Taking Eq. (\ref{heating_A}) at face value, we have then a simple power-law of the form 
\begin{equation}
\label{powerlaw1_A}
T_{\rm d}(a) =  T_0 \left({a\over a_0}\right)^{-1/(4+\beta)}. 
\end{equation}
Our aim is to arrive at a temperature distribution and for that we need some information about the GSD. A natural ansatz is the canonical MRN distribution \citep{Mathis77}, $n(a) \propto a^{-3.5}$, for which the GSD in terms of mass is $\varphi(a) \propto a^{-0.5}$. We thus have
\begin{equation}
\label{powerlaw2_A}
W(T_{\rm d}) = {dM_{\rm d}\over dT_{\rm d}} = {dM_{\rm d}\over da}\left|{dT_{\rm d}\over da}\right|^{-1} \propto  T_{\rm d}^{-3-\beta/2}.
\end{equation}
In this special case (thermal equilibrium, MRN distribution, grey-absorption limit) the temperature distribution is simply also a power-law, which serves well as a first approximation of the functional form of $W(T_{\rm d})$ since $\mathcal{J}_{\star}$ is typically only weakly dependent on the grain radius $a$.  A similar derivation of the power-law above can be found in \citet{Li99}.

In a more detailed picture the grey-absorption limit may not be strictly applicable and the slope of the temperature distribution is therefore likely steeper than in Eq. (\ref{powerlaw2_A}) and may also deviate from the simple power-law form above, i.e., $T_0$ would in such case be a function of $a$. However, regardless of whether $\mathcal{J}_{\star} = J_\star$ is a constant or not, Eq. (\ref{powerlaw2_A}) tells us that we must have $\varphi\to 0$ as $dT_{\rm d}/da \to 0$.  That is, absence of a temperature-size relation is not compatible with a the existence of grains having a range of sizes and temperatures, unless both heating and cooling take place at long-wavelengths, in which case the grain radius $a$ is cancelled out in Eq. (\ref{ebalance}).  
%This means we can also make the following (rather obvious) statement.\\[1mm]
%II. {\it For dust grains kept in thermal equilibrium by an invariant radiation field, the grain temperature is uncorrelated with grain size only if all grains have the same size and temperature, or if both heating and emission happens in the Rayleigh limit.}\\[1mm]
%In other words, the above means that as soon as there exist a grain-size (or temperature) distribution, then $dT_{\rm d}/da \neq 0$, because it is physically very implausible that dust grains would be heated by long-wavelength radiation. 

\section{Collisional heating}
{
\label{collheat}
In case heating is due mainly to collisions with ambient gas particles, we may write \citep{Dwek81,Dwek86,Dwek87}
\begin{equation}
\left({dE\over dt}\right)_{\rm coll} = \left({32 \pi\over m_{\rm e}}\right)^{1/2} a^2 n_{\rm e}\, (kT_{\rm e})^{3/2}\,h(a,T_{\rm e}), 
\end{equation}
where $k$ is Boltzmann's constant, $m_{\rm e}$ is the electron mass, $n_{\rm e}$ the electron density,  $T_{\rm e}$ the electron temperature and $h(a,T_{\rm e})$ is a unitless function describing the efficiency of energy deposition. By definition, $h = 1$ when the efficiency is maximal \citep[see][]{Dwek81}.  The fact that the heating rate is inversely proportional to the particle mass in the expression above, explains why collisions with electrons should be more important than collisions with any other gas particle, because the electron mass is very small in comparison with, e.g., the proton mass. Since the cooling rate is the same as in the case of radiative heating, i.e., cooling is still due to long-wavelength radiation, and adopting once again a power-law approximation for $Q'_{\rm abs}$ at long wavelengths, we obtain
\begin{equation}
\label{heating2}
\left({2\,k^3 T_{\rm e}^3 \over \pi^3 m_{\rm e}}\right)^{1/2}n_{\rm e}\,h(a,T_{\rm e})  \approx  {a\, Q'_0 \lambda_0^{\beta}} \tilde{\sigma}_{\rm SB}T_{\rm d}^{4+\beta},
\end{equation}
where all quantities are as previously defined. With $h\approx 1$ (efficient energy deposition), we recover a power-law of the same form as in the case of radiative heating. Thus, a power-law temperature distribution is a reasonable first approximation not only in case the dust heating is due to short-wavelength radiation, but also when it is due to efficient collisional heating. Another important aspect of the above is that it is difficult to have a dust population where all grains have very similar temperatures unless they also have very similar sizes, regardless of whether the heating is radiative or collisional.
}
\end{document}